\documentclass[sigconf]{acmart}

\usepackage{booktabs} 
\usepackage{subfig}
\usepackage{versions}
\usepackage{enumitem}
\usepackage{multirow}
\usepackage{amsmath,amssymb,amsfonts}
\usepackage{hyperref}
\usepackage{wrapfig}
\usepackage{lipsum}

\hypersetup{
  colorlinks = true,
  linkcolor = {blue},
  citecolor = {cyan},
}

\usepackage{graphicx}

\author{Shuai Li}
\affiliation{%
  \institution{University of Minnesota}
  \city{Minneapolis}
  \state{MN}
}
\email{lixx2381@umn.edu}

\author{Huajun Guo}
\affiliation{%
  \institution{University of Minnesota}
  \city{Minneapolis}
  \state{MN}
}
\email{guoxx663@umn.edu}

\author{Nicholas Hopper}
\affiliation{%
  \institution{University of Minnesota}
  \city{Minneapolis}
  \state{MN}
}
\email{hoppernj@umn.edu}

\begin{document}

\copyrightyear{2018}
\acmYear{2018}
\setcopyright{acmcopyright}
\acmConference[CCS '18]{2018 ACM SIGSAC Conference on Computer and Communications Security}{October 15--19, 2018}{Toronto, ON, Canada}
\acmBooktitle{2018 ACM SIGSAC Conference on Computer and Communications Security (CCS '18), October 15--19, 2018, Toronto, ON, Canada}
\acmPrice{15.00}
\acmDOI{10.1145/3243734.3243832}
\acmISBN{978-1-4503-5693-0/18/10}

\title{Measuring Information Leakage in Website Fingerprinting Attacks and Defenses} 

\begin{abstract}

Tor provides low-latency anonymous
and uncensored network access against a local or network adversary. Due to
the design choice to minimize traffic overhead (and increase the pool of
potential users) Tor allows some information about the client's connections to
leak. Attacks using (features extracted from)
this information to infer the website a user visits are called Website
Fingerprinting (WF) attacks. We develop a methodology and tools to
measure the amount of leaked information about a website. We apply this tool to a comprehensive set of features extracted from a large set of websites and WF defense mechanisms, allowing
us to make more fine-grained observations about WF attacks and defenses.
\end{abstract}



\maketitle


\section{Introduction}

The Tor anonymity network uses layered encryption and traffic relays
to provide private, uncensored network access to millions of users
per day.  This use of encryption hides the exact contents of messages
sent over Tor, and the use of sequences of three relays prevents any single
relay from knowing the network identity of both the client and the
server.  In combination, these mechanisms provide effective resistance
to basic traffic analysis.

However, because Tor provides low-latency, low-overhead communication, it does
not hide traffic features such as the volume, timing, and direction of
communications. Recent works~\cite{wang2014effective,ndss16,kfingerprint} have shown that these
features leak information about which website has been visited to the extent
that a passive adversary that records this information is able to train a classifier to recognize the
website with more than $90\%$ accuracy in a closed-world scenario with $100$ websites. This attack is often referred to as a Website
Fingerprinting (WF) attack. 
In response, many works~\cite{wang2014effective,buflo,cai2014,glove,decoypage,morphing,pipelining,wpes14-csbuflo,cherubin2017website,wang2015walkie} have proposed defenses that attempt to hide this
information, by padding connections with extra traffic or rearranging the
sequence in which files are requested. 

\medskip


\noindent \textbf{Defense Evaluation.} To evaluate a defense, the popular practice is to train classifiers based on altered traffic characteristics and evaluate the effect of the defense by classification accuracy. If the accuracy of the classifier is low enough, the defense is believed to be secure with minimal information leakage; one defense is believed to be better than another if it results in lower accuracy.

\begin{figure} [b]
\centering
\includegraphics[width=0.32\textwidth]{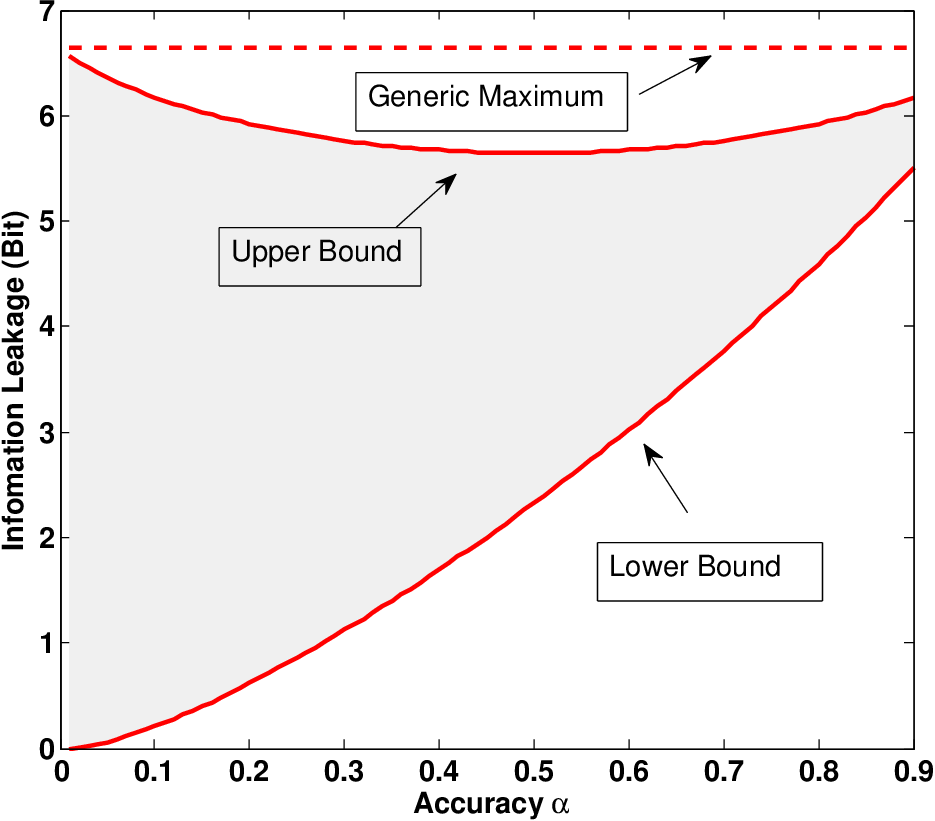}
    \caption{Accuracy vs. Information Leakage. {\normalfont This figure shows the range of potential information leakage for a given classification accuracy (in the closed-world setting with 100 websites)}}\label{fig:acc_vs_inforange}
\end{figure}

{\em Accuracy vs. Information Leakage.} We raise a question: does low accuracy always mean low information leakage from WF defenses? Our answer is no. The first reason is that accuracy is classifier-dependent. It is possible that the information leakage of a WF defense is high, but the classifier is ineffective, so that its accuracy is low. More importantly, 
accuracy is all-or-nothing: classifiers output a single guess and if it is wrong, this is judged to mean the defense has been successful. But it ignores cases where a classifier may confuse some pages with a small set of others. In such situations, an attacker may well be able to significantly reduce the set of likely pages represented by a fingerprint, even if they cannot reliably choose the correct page from among this set. We can see that the fingerprint can contain a great deal of {\em information} about the web page even if the classifier cannot accurately identify the correct page. Accuracy is prone to {\em underestimate} the information leakage in WF defenses, or in other words, low accuracy doesn't necessarily mean low information leakage.

We further prove the above observation by the information-theoretic quantification upon a given accuracy. We find that in a closed-world setting with $n$ websites, a feature set yielding a classifier with accuracy $\alpha$ could leak information through the classifier with the uncertain range $(1-\alpha)\log_2 (n-1)$ (the difference between the maximum and minimum). The proof also shows that such uncertainty increases with lower accuracy. Figure \ref{fig:acc_vs_inforange} shows that when $n=100$
and $\alpha = 0.95$, the uncertain range is only $0.33$ bit; but when $\alpha = 0.05$, the possible leakage could be as high as 6.36 bits and as low as 0.06 bits! This uncertainty reveals the potential discrepancy between information leakage and accuracy in evaluating WF defenses, though its impact on WF attacks is limited. Low information leakage implies low classification accuracy, but the converse is not necessarily true, thus we argue that \textbf{validating WF defenses by accuracy alone is flawed}.



\medskip

\noindent \textbf{Feature Evaluation.} Different features may carry different amounts of information. WF defense designers can evaluate features to find more informative ones to hide~\cite{cai2014}; attackers can do so to discover highly informative features and optimize their feature set~\cite{kfingerprint}. Existing works~\cite{cai2014,kfingerprint} designed comparative methods to rank the features by their information leakage, but these methodologies do not give a straightforward way to quantify the {\em relationships} between features. How much information do features A and B share? If feature A is more informative than features B or C alone, are features B and C together more informative than A? These methodologies are unable to answer these questions.

\medskip

We argue that these coarse-grained evaluations of features and defenses are overly simplistic. The analysis of new WF attack features and defenses should start with the question: how much information is leaked? To answer this question, two challenges should be addressed. The first challenge is finding a way to model the behavior of WF features and the interaction between them; these features can have highly complex relationships and behavior, exhibiting distributions that could be discrete, continuous, or even partly discrete and partly continuous. The second challenge is the curse of dimensionality when estimating the total information leakage, as the state-of-art feature sets are usually high-dimensional. Unfortunately, existing works~\cite{cav,mather2012pinpointing} limited their experimental measurement to features' {\em individual} information leakage, and they cannot overcome these challenges.   

\medskip

\noindent \textbf{Information Leakage Measurement Framework.}
In this paper, we develop WeFDE (for \textbf{We}bsite \textbf{F}ingerprint \textbf{D}ensity \textbf{E}stimation), a methodology for modelling the likelihood functions of website fingerprints, and a set of tools for measuring the fingerprints' information leakage. To address the first challenge, WeFDE uses adaptive kernel density estimation~\cite{van2003adaptive} to model the probability density function of a feature or a category of features. By allowing kernels to determine their bandwidth separately, we adaptively model both continuous and discrete density functions; by estimating multi-dimensional kernels over sets of features, we model the interactions between features. We address the second challenge by introducing a set of dimension reduction approaches. Firstly, we measure features' pairwise mutual information to exclude redundant features. Secondly, we use Kononenko's Algorithm~\cite{cheng1999comparing,Kononenko} and DBSCAN~\cite{ester1996density} to separate features into sub-groups, which have pairwise mutual information higher than a threshold $\epsilon$ within each group, and lower than $\epsilon$ among different groups. Then we apply adaptive kernels for each sub-group with reduced dimensionality. Finally, our experiment shows that by including enough highly informative features we are able to approximate the overall information leakage. This enables us to further reduce the dimensionality of our measurement.

\medskip

\noindent \textbf{Measurement Results.}
We apply WeFDE to a comprehensive list of $3043$ features (including, to the best of our knowledge, all features in the Tor WF literature~\cite{shi2009fingerprinting,kfingerprint,wpes11-panchenko,buflo,ndss16,wang_wpes,cai1,wang2014effective}) extracted from a $211219$ Tor web browsing visits for about $2200$ websites. Among the features of WF attacks, we find that: (a) $45.36\%$ of $183$ most informative features are redundant; (b) an individual feature leaks no more than $3.45$ bits
information in the closed-world setting with $100$ websites, which is the maximum leakage we observe in our experiment from the feature of rounded outgoing packet count; (c) download stream, though having more packets than upload stream, leaks less information; (d) a larger world size has little impact on a WF feature's individual information leakage. We also include WF defenses such as Tamaraw~\cite{cai2014}, BuFLO~\cite{buflo}, Supersequence~\cite{wang2014effective}, WTF-PAD~\cite{juarez2016toward}, and CS-BuFLO~\cite{wpes14-csbuflo} to study the discrepancy between accuracy and information leakage. Our experimental results confirm this discrepancy and demonstrate that accuracy alone is not reliable to validate a WF defense or compare multiple ones. We also find that the information leakage of WTF-PAD~\cite{juarez2016toward} is unusually high. Interestingly, recent work~\cite{2018arXiv180102265S} confirms our result by achieving $90\%$ classification accuracy against WTF-PAD.

\medskip

\noindent \textbf{Contributions.} 
We provide our contributions as follows.
First, this paper identifies that validating WF defenses by accuracy alone is flawed. By information-theoretic quantification, we find that when accuracy is low, its corresponding information leakage is far from certain. Second, we propose WeFDE which makes it possible to measure the joint information leakage from a large set of features. In contrast, existing works only limited their experimental measurement to features' individual information leakage, and they
cannot cope with features of complex property. WeFDE overcomes these two limitations. We also release the source code of WeFDE in the GitHub repository\footnote{https://github.com/s0irrlor7m/InfoLeakWebsiteFingerprint}. Third, we use WeFDE to perform information leakage measurement for all $3043$ features proposed in the Tor website fingerprinting literature, based on a large dataset having $211219$ Tor web browsing visits to about $2200$ websites. As far as we know, our work is the first large-scale information leakage measurement in the literature. Fourth, our measurement results provide the new information-theoretic
insights upon WF features, and these results give the empirical confirmation that accuracy is not reliable to validate a WF defense or compare multiple ones.

The paper is organized as follows: Section~\ref{sec:bg} and Section~\ref{sec:related} give background and related works on WF attacks and defenses, with Section~\ref{sec:features} introducing the
features.
Section~\ref{sec:system} introduces the system design of WeFDE.
Section~\ref{sec:closed}, Section~\ref{sec:validation}, Section~\ref{sec:info_acc}, and Section~\ref{sec:open} give information leakage measurement results, and Section~\ref{sec:discuss} provides discussions. Finally, we conclude in Section~\ref{sec:conclusion}.

\vspace{-6pt}
\section{WF Attack Models}\label{sec:bg}

A WF attacker aims at learning which website a user has visited. The attacker can be an Internet Service Provider (ISP) or a malicious Tor entry guard. It is supposed to be passive (no packet manipulation), but it can eavesdrop the traffic originated from or destinated to the user. Without turning to traffic contents or its IP addresses (both can be encrypted or obfuscated), the attacker inspects the traffic fingerprints for detection. These fingerprints can be packet length or the transmission time. Neither Cryptographic algorithms nor the anonymous services such as Tor can cover such fingerprints. State of art attacks~\cite{kfingerprint,wang2014effective,ndss16} demonstrate that the fingerprints carry sufficient information that the attacker can pinpoint the visited website by more than $90\%$ accuracy (with assumptions). In the following, we introduce two attack models of the website fingerprinting attack.

\textbf{Closed-World Attack Model.}  
An attacker in the closed-world knows a set of websites $C = \{c_1, c_2, \cdots, c_n\}$ the user may visit. We adopt an equal-prior model, in which the user visits a website with probability $1/n$. The attacker's goal is to decide which oneis visited.

\textbf{Open-World Attack Model.}
The attacker in this attack model has a set of websites for monitoring; its goal is to decide whether the user visited a monitored website or not, and if yes, which monitored website. Though the user may visit any website, a non-monitored set of websites are introduced to approximate the user visiting the non-monitored websites. We consider a popularity-prior model, in which we give prior probabilities to websites by their popularity, without considering whether the websites are monitored or not.

\vspace{-6pt}
\section{Related Work} \label{sec:related}

\textbf{Website Fingerprinting Attacks.} The first category of WF attacks targeted encrypted protocols with no packet length hiding~\cite{2006}. 
More recent website fingerprinting attacks focus on Tor anonymous service, in which the unique packet length is hidden by fixed-size Tor cells. Cai {\em et al.}~\cite{cai1} used edit distance to compare Tor packet sequences, and achieved $86\%$ accuracy in the closed-world scenario with $100$ websites. Wang and Goldberg~\cite{wang_wpes} further improve the accuracy to $91\%$ by using Tor cells instead of TCP/IP packets, deleting SENDMEs, and applying new metrics such as fast Levenshtein. Their later work~\cite{wang2014effective} increases the accuracy by using a KNN classifier. Panchenko {\em et al.}~\cite{ndss16} introduces a new method to extract the packet number information, which increases the accuracy by $2\%$. Recently, Hayes and Danezis~\cite{kfingerprint} use random forests to construct the current state-of-art website fingerprinting attack. 

\textbf{Website Fingerprinting Defenses.} Several defenses have been proposed to defeat WF attacks. One category of defenses try to randomize the traffic fingerprints by traffic morphing~\cite{morphing}, loading a background page~\cite{decoypage}, or randomized pipelining~\cite{pipelining}. These are demonstrated ineffective by several works~\cite{cai1,wang2014effective}.  

Another category of defenses try to hide traffic features by deterministic approaches. By holding the packets or creating dummy packets, BuFLO~\cite{buflo} requires the packets sent in fixed size and fixed time interval. The packets are padded until reaching a transmission time threshold $\tau$ if their original transmission time is shorter. Otherwise, BuFLO lets the traffic finish. CS-BuFLO~\cite{wpes14-csbuflo} is proposed to extend BuFLO to include congestion sensitivity and some rate adaptation. Tamaraw~\cite{cai2014} improves the efficiency of BuFLO by two methods. First, it allows different transmission rate for outbound and inbound traffic. Second, it pads to make the packet count a multiple of parameter $L$: if the packet number in one direction is more than $nL$ and less than $(n+1)L$, it sends padding packets until the count is $(n+1)L$. Supersequence~\cite{wang2014effective} utilizes clustering algorithms to group websites. For each group of websites, Supersequence computes a super trace to be the manner of transmitting the instances of the websites under this group. WTF-PAD~\cite{juarez2016toward} uses adaptive padding to be efficient.
Our paper includes these defenses for information leakage evaluation. We leave recently proposed defenses~\cite{cherubin2017website,wang2015walkie} in our future work.

\textbf{Website Fingerprinting Evaluation.} Juarez {\em et al.}~\cite{critical} evaluates the effectiveness of WF attacks in practical scenarios, enumerating several assumptions about user settings, adversary capabilities, and the nature of the web that do not always hold. Without these assumptions, the accuracy of the attacks are significantly decreased. Cai {\em et al.}~\cite{cai2014} use a comparative method to analyze defenses. They apply generators to transform a website class $C$ into $C'$, making $C$ and $C'$ differ only by one (category of) feature. Then they evaluate whether a specific defense is successful in hiding the feature. Though they claim this method can shed light on which features convey more information, the information leakage comparison between features is unclear and not quantified.

Cherubin~\cite{naive_not_bayes} provides the lower bound estimate for an attacker's error. The error of the Nearest Neighbor classifier is used to estimate the lower bound of the Bayes error, which is further used to be the lower bound for the error of any classifier. Based on such lower bound, a new privacy metric called $(\xi, \Phi)$--privacy is proposed. Though this privacy metric is not dependent on any specific classifier, it is still a variant of the accuracy/error metric, and therefore the flaw of accuracy also applies to $(\xi, \Phi)$--privacy.

\textbf{Security Measurement by Information Theory.} 
There are two information leakage studies in website fingerprinting attacks, but they are limited from quantification methodology and dataset size, to feature set and analysis. Such limitations prevent them from answering the question: how much information in total is leaked from the website fingerprint?

Chothia {\em et al.}~\cite{cav} proposed a system called leakiEst to quantify information leakage from an observable output about a single secret value. Then leakiEst was applied on e-passports and Tor traffic. However, leakiEst can only measure the information leakage of a {\em single} feature, rather than a category of features or all features. In addition, it only included $10$ visits for $500$ websites in their dataset, and it just considered $75$ features. Furthermore, leakiEst cannot deal with information leakage under various scenarios, such as open-world setting and setting with defenses. 

Mather {\em et al.}~\cite{mather2012pinpointing} quantified the side-channel information leakage about the user's inputs in the web applications. However, the work shared many of the above limitations. The experiment only considered packet size features; the size of the dataset was unknown; the quantification only came to a {\em single} feature. Though the work included the multivariate extension, it didn't apply it in the experiment, and it didn't have dimension reduction to handle the curse of the dimensionality.


In comparison, our paper overcomes these limitations, and it can measure information leakage from a category of features. Specifically, it includes all $3043$ features proposed in Tor WF literatures and much larger dataset with $211219$ visits. The resulting information leakage quantification is therefore more accurate and representative. It's able to quantify information leakage of a set of features with high dimension, so that it can tell how much information is leaked in total. 
More importantly, our paper not only quantifies the information leakage, but also reveals and analyzes the discrepancy between accuracy and information leakage when validating WF defenses. Our exprimental results demonstrate the flaw of accuracy.




\textbf{Mutual Information Estimation.}
Kernel Density Estimate (KDE) and k-nearest neighbors (k-NN) are two popular approaches to estimate the mutual information between two random variables. They are believed to outperform other estimators due to their ability to capture the nonlinear dependence where present~\cite{kde_knn}. In this paper, we choose KDE instead of k-NN for the following reasons. Firstly, k-NN tends to underestimate mutual information between strongly dependent variables~\cite{strong1,strong2}. In security settings, this means the measured information leakage would be less than it should be, making k-NN unsuitable for information leakage measurement. More importantly, we find that the KSG estimator~\cite{ksg}, which is the most popular k-NN estimator, is unable to handle a categorial random variable. This limitation matters because in Section \ref{sec:method} we will see one of variables is about website information, which is categorial. We further confirm the above two reasons by an experiment. We find the total information leakage measured by the KSG estimator is around $2$ bits in a closed-world setting with 100 websites, much lower than what WF attacks have shown. Thus, we choose the KDE approach in our paper.

\begin{table}\small 
\begin{tabular}{ c c c c c }
  \hline
  && Source & Instances & Batches \\
  \hline
  \hline
  &Closed-World &Alexa 1-100 & 55779 & 20 \\
   
  \parbox[t]{6mm}{\multirow{2}{1cm}{Open-World}} &Monitor & \cite{wang2014effective} &  17985 & 8 \\
   &Non-Monitor & Alexa 1-2000 & 137455 & 10\\
  \hline 
\end{tabular}
    \caption{DATASET. {\normalfont We adopt Crawler \cite{tor-browser-crawler} to collect the network traffic in batches. This crawler uses Selenium to automate the Tor Browser Bundle and applies Stem library to control the Tor process. It extended the circuit renewal period to 600,000 minutes and disabled \texttt{UseEntryGuard} to avoid using a fixed set of entry guards. We apply the method in \cite{ndss16} to extract the cell packets from the traffic and measure the information leakage based
    on the cell packets.}\protect\footnotemark }\label{tab:dataset}
\vspace{-10pt}
\end{table}
\footnotetext{Our dataset allows the websites to have different number of instances. This uneven distribution is mostly caused by the failed visits in the crawling process. Note that it doesn't impact our information leakage measurement.}

\section{Traffic and its features} \label{sec:features}
A user's traffic is a sequence of packets with timestamps which are originated from or destinated to it. 
We use $T(C)$ to denote the traffic when the user visited the website $C$. Then\begin{equation}
T(C) = \langle (t_0, l_0), (t_1, l_1), \cdots, (t_m, l_m) \rangle\ ,
\end{equation}
where $(t_i, l_i)$ corresponds to a packet of length $|l_i|$ in bytes with a timestamp $t_i$ in seconds. The sign of $l_i$ indicates the direction of the packet: a positive value denotes that it is originated from the server, otherwise the user sent the packet. Table~\ref{tab:dataset} describes our collected traffic for information leakage measurement.

In the state-of-art website fingerprinting attacks~\cite{ndss16,kfingerprint, wang2014effective}, it is the features of the traffic rather than the traffic itself that an attacker uses for deanonymization. One of the contribution of this paper is that it measures a {\em complete} set of existing traffic features in literatures of website fingerprinting attacks in Tor~\cite{shi2009fingerprinting,kfingerprint,wpes11-panchenko,buflo,ndss16,wang2014effective}. Table \ref{tab:feature_set} summarizes these features by category. More details about the feature set can be found in Appendix \ref{sec:appfeature}.

\begin{table}\small 
\centering
\begin{tabular}{ | c | l | c |}
  \hline
    Index & Category Name [Adopted by] & No.\\
    \hline
    \hline
   1 & Packet Count~\cite{kfingerprint,wpes11-panchenko,buflo,wang2014effective,ndss16} & 13\\
   2 & Time Statistics~\cite{kfingerprint,buflo,wang2014effective} & 24\\
   3 & Ngram [this paper] & 124 \\
   4 & Transposition~\cite{kfingerprint,wang2014effective} & 604\\
   5 & Interval-I~\cite{kfingerprint,wang2014effective} & 600\\
   6 & Interval-II~\cite{shi2009fingerprinting} & 602\\
   7 & Interval-III~\cite{wpes11-panchenko} & 586\\
   8 & Packet Distribution~\cite{kfingerprint} & 225\\
   9 & Bursts~\cite{wang2014effective} & 11\\
   10 & First 20 Packets~\cite{wang2014effective} & 20\\
   11 & First 30 Packets~\cite{kfingerprint} & 2\\
   12 & Last 30 Packets~\cite{kfingerprint} & 2\\
   13 & Packet Count per Second~\cite{kfingerprint} & 126\\ 
   14 & CUMUL Features~\cite{ndss16} & 104\\
  \hline
\end{tabular}
\caption{Feature Set. {\normalfont $3043$ features from $14$ categories}}\label{tab:feature_set}
\vspace{-10pt}
\end{table}

\vspace{-6pt}
\section{System Design}\label{sec:system}

\subsection{Methodology}\label{sec:method}
\begin{figure}[b]
  \centering
  \includegraphics[width=0.4\textwidth]{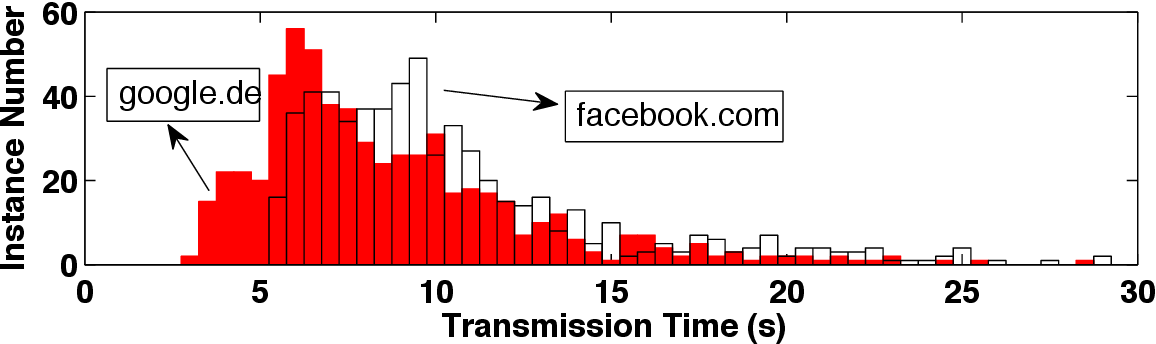}
  \includegraphics[width=0.41\textwidth]{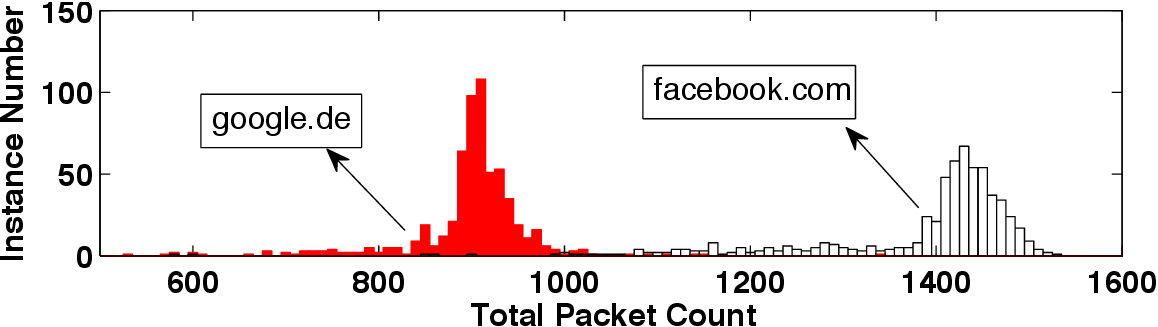}
\vspace{-5pt}
  \caption{Different features may carry different amount of information. {\normalfont Take transmission time and total packet count as an example. This figure shows the latter carries more information in telling which website is visited (www.google.de or www.facebook.com)}}
  \label{fig:info_leak}
\vspace{-10pt}
\end{figure}

The features leak information about which website is visited. Total packet count is a good example. Figure \ref{fig:info_leak} shows that visiting \texttt{www.google.de} creates 700 to 1000 packets, while browsing \texttt{www.facebook.com} results in 1100 to 1600 packets. Suppose an attacker passively monitors a Tor user's traffic, and it knows that the user has visited one of these two websites (closed-world assumption). By inspecting the total packet count of the traffic, the attacker can tell which website is visited. 

Different features may carry different amounts of information. Figure \ref{fig:info_leak} displays the download time in visiting \texttt{www.google.de} and \texttt{www.facebook.com}. The former loads in about 3 to 20 seconds, and the latter takes 5 to 20 seconds; Their distributions of download time are not easily separable. As a result, the attacker learns much less information from the download time than from total packet count in the same closed-world scenario.

This raises question of how to quantify the information leakage for different features. We adopt mutual information~\cite{info_book}, which evaluates the amount of information about a random variable obtained through another variable, which is defined as:

\medskip

\noindent\textbf{DEFINITION.} Let $F$ be a random variable denoting the traffic's fingerprint, and suppose $W$ to be the website information, then $I(F;W)$ is the amount of information that an attacker can learn from $F$ about $W$, which equals to:
\begin{equation}
I(F;W) = H(W) - H(W|F)
\end{equation}
$I(\cdot)$ is mutual information, and $H(\cdot)$ is entropy. In the following, we describe our system to measure this information leakage.

\vspace{-6pt}
\subsection{System Overview}

Aimed at quantifying the information leakage of a feature or a set of features, we design and develop our \textbf{We}bsite \textbf{F}ingerprint \textbf{D}ensity \textbf{E}stimation, or WeFDE. Compared with existing systems such as leakiEst~\cite{cav}, WeFDE is able to measure joint information leakage for more than one feature, and it is particularly designed for measuring the leakage from WF defenses, in which a feature could be partly continuous and partly discrete.

\begin{figure}[t]
  \centering
  \includegraphics[width=0.42\textwidth]{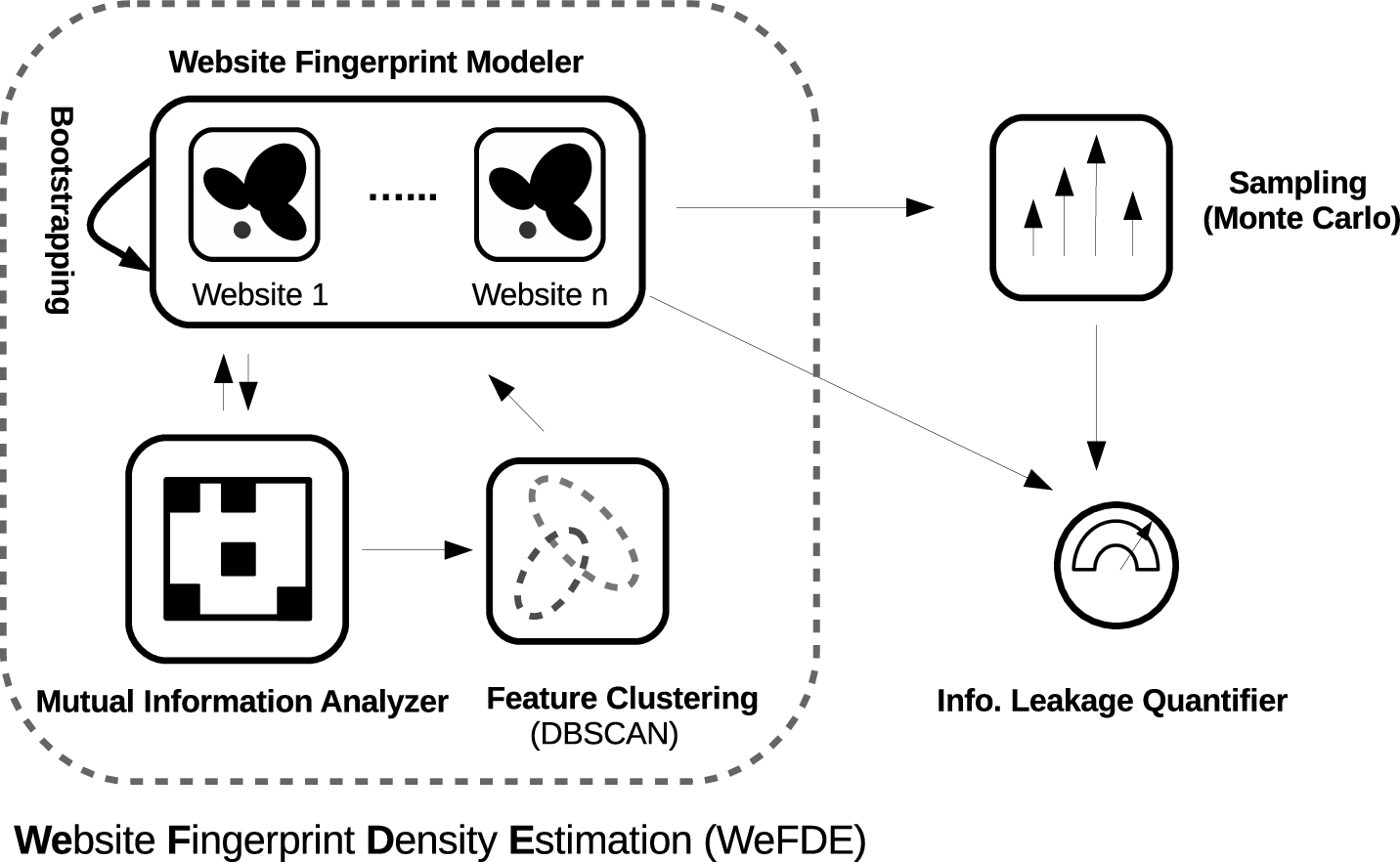}
  \vspace{-5pt}
  \caption{WeFDE's Architecture}
  \label{fig:arch}
  \vspace{-10pt}
\end{figure}

Figure~\ref{fig:arch} shows the architecture of WeFDE. The information leakage quantification begins with the Website Fingerprint Modeler, which estimates the probability density functions of features. In case of measuring joint information of features, Mutual Information Analyzer is activated to help the Modeler to refine its models to mitigate the curse of dimensionality. During the information leakage quantification, the Website Fingerprint Modeler is used to generate samples. By
Monte Carlo approach~\cite{importance_sampling} (see Appendix \ref{sec:mc} for more information), the Information Leakage Quantifier derives the final information leakage by evaluating and averaging the samples' leakage. In the following, we describe our modules.

\vspace{-6pt}
\subsection{Website Fingerprint Modeler}

The task of Website Fingerprint Modeler is to model the probability density function (PDF) of features. A popular approach is to use a histogram. However, as the traffic features exhibit a great range of variety, it's hard to decide on the number of bins and width. WeFDE adopts Adaptive Kernel Density Estimate (AKDE)~\cite{rosenblatt1956remarks}, which outperforms histogram in smoothness and continuity. AKDE is a non-parametric method to estimate a random variable's PDF. It uses kernel functions---a non-negative function that integrates to one and has mean zero---to approximate the shape of the distribution.

Choosing proper bandwidths is important for AKDE to make an accurate estimate. WeFDE uses the plug-in estimator~\cite{bw_survey} for continuous features, and in case of failure, WeFDE uses the rule-of-thumb approach~\cite{bw_survey} as the alternative. If the feature is discrete, we let the bandwidth be a very small constant (0.001 in this paper). The choice of the small constant has no impact on the measurement, as long as each website uses the same constant as the bandwidth. 

To model the features' PDFs in WF defenses, WeFDE has two special properties. Firstly, our AKDE can handle a feature which is partly continuous and partly discrete (or in other words, a mixture of continuous and discrete random variables). Such features exist in a WF defense such as BuFLO~\cite{buflo} which always sends at least T seconds. These features would be discrete if the genuine traffic can be completed within time $T$, otherwise, the features would be continuous. Secondly, our AKDE is able to distinguish a continuous-like discrete feature. Take transmission time as an example. This feature is used to be continuous, but when defenses such as Tamaraw~\cite{cai2014} are applied, the feature would become discrete. Our modeler is able to recognize such features. For more details, please refer to Appendix \ref{sec:kde}. 

We further extend WeFDE to model a set of features by adopting the multivariate form of AKDE. However, when applying multivariate AKDE to estimate a high dimensional PDF, we find AKDE inaccurate. The cause is the curse of dimensionality: as the dimension of the PDF increases, AKDE requires exponentially more observations for accurate estimate. Considering that the set of features to be measured jointly could be large (3043 features in case of total information measurement), we need dimension reduction techniques. In the following, we introduce our Mutual Information Analyzer to mitigate the curse of dimensionality.


\vspace{-6pt}
\subsection{Mutual Information Analyzer}

The introduction of Mutual Information Analyzer is for mitigating the curse of dimensionality in multivariate AKDE. It helps the Website Fingerprint Modeler to prune the features which share redundant information with other features, and to cluster features by dependency for separate modelling.

This Analyzer is based on the features' pairwise mutual information. 
To make the mutual information of any two features have the same range, WeFDE normalizes it by Kvalseth's method~\cite{kvalseth1987entropy} (other normalization approaches~\cite{normalize} may also work). Let $\mathrm{NMI_{max}}(c,r)$ denote the normalized mutual information between feature $c$ and $r$, then it equals to:
\begin{equation*}
\mathrm{NMI_{max}}(c,r) = \frac{I(c;r)}{\max\{H(c),
  H(r)\}} 
\end{equation*}         
\medskip
Since $I(c;r)$ is less than or equal to $H(c)$
and $H(r)$, $\mathrm{NMI_{max}}(c,r)$ is in $[0,1]$. A higher value of $\mathrm{NMI_{max}}(c,r)$ indicates higher dependence between $r$ and $c$, or in other words, they share more information with each other.


\medskip

\noindent{\bf Grouping By Dependency.}
A workaround from curse of dimensionality in higher dimension is to adopt Naive Bayes method, which assumes the set of features to be measured is conditionally independent. Naive Bayes requires many fewer observations, thanks to the features' probability distribution separately estimated. However, we find dependence between some features of the website fingerprint, violating the assumption of Naive Bayes. 

We adopt Kononenko's algorithm (KA)~\cite{cheng1999comparing,Kononenko}, which clusters the highly-dependent features into disjoint  
groups. In each group, we model the joint PDF of its features by applying AKDE. Among different groups, conditional independence is assumed. KA takes the advantage of how Naive Bayes mitigates the curse of dimensionality, while keeping realistic assumptions about conditional independence between groups.

We use clustering algorithms to partition the features into disjoint groups. An ideal clustering algorithm is expected to guarantee that any two features in the same group have dependence larger than a threshold, and the dependence of the features in different groups is smaller than the same threshold. This threshold allows us to adjust independence degree between any two groups. We find that DBSCAN~\cite{ester1996density} is able to do so.

DBSCAN is a density-based clustering algorithm. It assigns a feature to a cluster if this feature's distance from any feature of the cluster is smaller than a threshold $\epsilon$, otherwise the feature starts a new cluster. Such a design enables DBSCAN to meet our goal above. To measure features' dependence, we calculate their normalized pairwise mutual information matrix $M$; then to fit in with DBSCAN, we convert $M$ into a distance matrix $D$ by $D = \mathbf{1} - M$, where $\mathbf{1}$ is a matrix of ones. A feature would have distance $0$ with itself, and distance $1$ to an independent feature. We can tune $\epsilon$ in DBSCAN to adjust the degree of independence between groups. We choose $\epsilon=0.4$ in the experiments based on its empirical performance in the trade-off between its impact on information measurement accuracy and KA's effectiveness in dimension reduction.

We model the PDF of the fingerprint by assuming independence between groups.
Suppose KA partitions the fingerprint $\vec{\mathbf{f}}$ into $k$ groups, $\vec{\mathbf{g}}_1, \vec{\mathbf{g}}_2, \cdots, \vec{\mathbf{g}}_k$, with each feature belonging to one and only one group. To evaluate the probability $p(\vec{\mathbf{f}}|c_j)$, we instead calculate $\hat{p}(\vec{\mathbf{g}}_1|c_j) \hat{p}(\vec{\mathbf{g}}_2|c_j) \cdots \hat{p}(\vec{\mathbf{g}}_k|c_j)$, where $\hat{p}(\cdot)$ is the PDF estimated by AKDE.  

As a hybrid of the AKDE and Naive Bayes, Kononenko's algorithm avoids
the disadvantages of each. First, Kononenko's algorithm does not have
the incorrect assumption that the fingerprint features are independent. It
only assumes independence between groups, as any two of them have
mutual information below $\epsilon$. Second, Kononenko's algorithm mitigates the curse of dimensionality. The groups in Kononenko's
algorithm have much less features than the total number of features.

\medskip

\noindent{\bf Dimension Reduction.}
Besides the KA method to mitigate the curse of dimensionality, we employ two other approaches to further reduce the dimension.

The first approach is to exclude features being represented by other features. We use the pairwise mutual information to find pairs of features that have higher mutual information than a threshold ($0.9$ in this paper). Then we prune the feature set by eliminating one of the features and keeping the other.  

\begin{figure*}[h]
\centering
\includegraphics[width=0.75\textwidth]{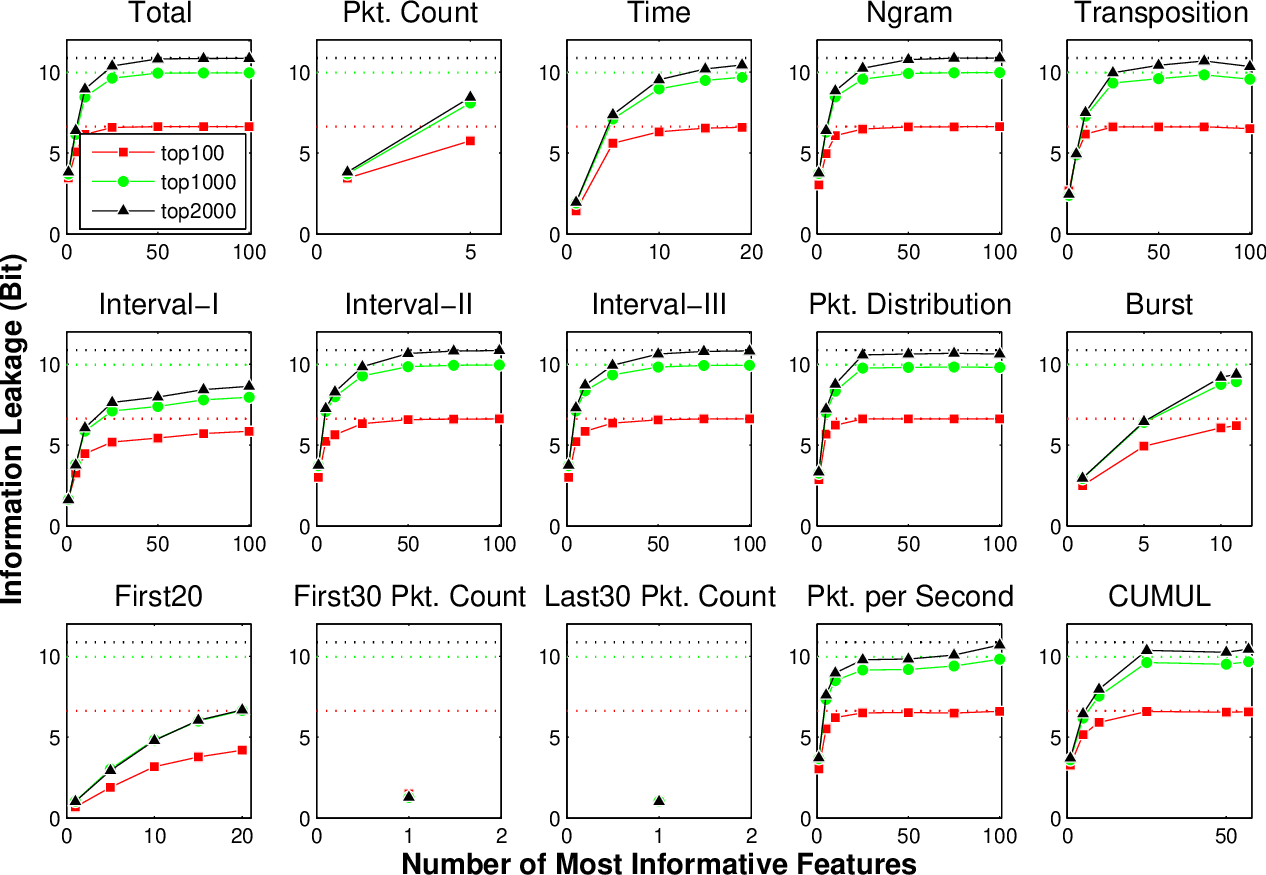}
\vspace{-5pt}
\caption{Closed-World Setting: Information Leakage by Categories (bit). {\normalfont The aggregate information from most categories of feature grows with the world size, indicating that these categories of features can be used to distinguish between larger sets of websites}}
\label{fig:closed_category}
\vspace{-10pt}
\end{figure*}

Our second approach is to pick out a number of the most informative features to approximate all features' information leakage. Given a set of features to measure, we sort the features by their individual information leakage. Instead of measuring all features' information leakage, we pick out top $n$ features that leak the most information about the visited websites. The measurement results by varying $n$ are shown in Figure~\ref{fig:closed_category} and Figure~\ref{fig:openw_category}. It shows that with $n$ increasing, the top $n$ features' information leakage would increase at first but finally reach a plateau. This phenomenon shows that the information leakage of sufficient top informative features is able to approximate that of the overall features. Such observation is also backed by~\cite{kfingerprint}, which discovered that including more top informative features beyond $100$ had little gain for classification.

\begin{figure}[b]
\centering
\vspace{-10pt}
\includegraphics[width=0.32\textwidth]{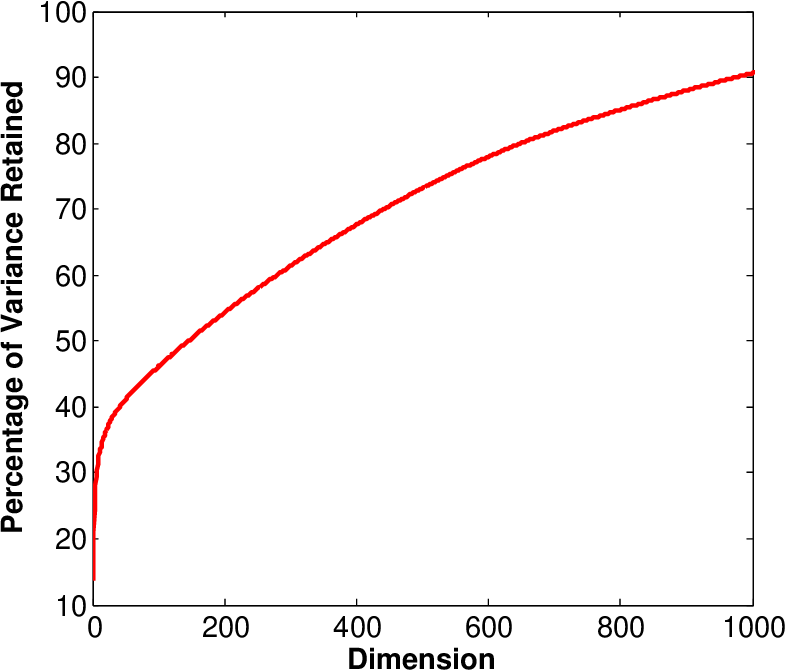}
\vspace{-10pt}
\caption{the Percentage of Variance Retained in PCA. {\normalfont Percentage of variance indicates the information loss in PCA}}\label{fig:PCA}
\vspace{-6pt}
\end{figure}

We didn't choose other dimension reduction methods such as Principal Component Analysis (PCA)~\cite{jolliffe1986principal}. Our goal is to mitigate the curse of dimensionality in modelling {\em website fingerprints} by AKDE; but methods like PCA transform website fingerprints into {\em opaque components} which are much less understandable. More importantly, our experimental results demonstrate the poor performance of PCA. Figure \ref{fig:PCA} shows that the percentage of variance retained when PCA reduces to a specific dimension. Note that the percentage of variance is the popular approach to estimate the information loss in PCA. It displays that if our goal is to reduce the dimension from $3043$ to $100$, the percentage of variance retained after PCA is under $50\%$, indicating high information loss. Thus, PCA doesn't fit in our case. 

\medskip

\begin{figure}[t]
  \centering
  \includegraphics[width=0.45\textwidth]{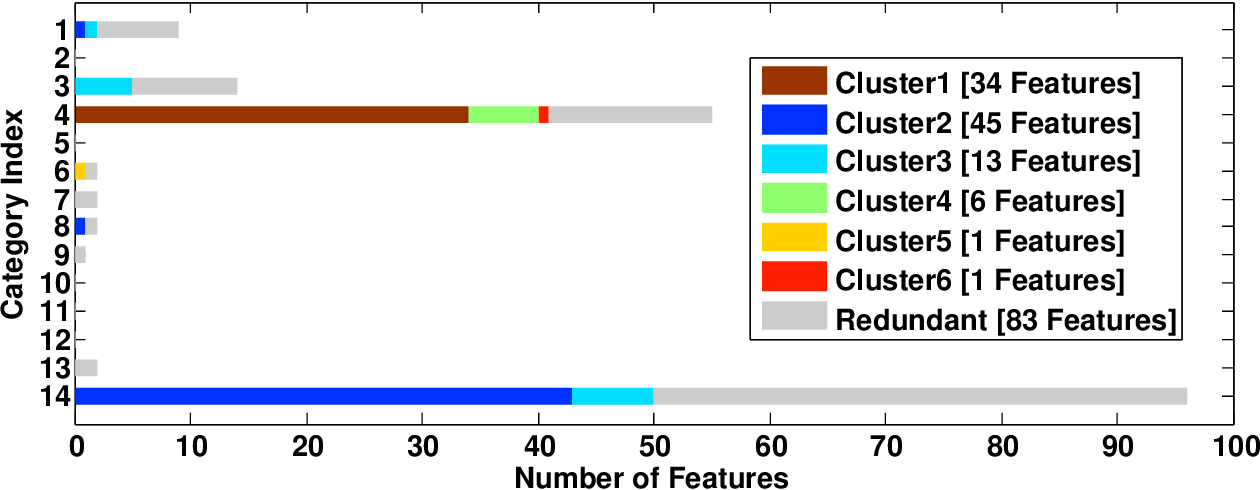}
  \vspace{-5pt}
  \caption{The Clustering Results of Mutual Information Analyzer. {\normalfont $6$ clusters are found for the most informative and non-redundant $100$ features.}}
  \label{fig:mi_by_category}
  \vspace{-10pt}
\end{figure}

\noindent\textbf{The Results.} Figure \ref{fig:mi_by_category} displays the outcome of our Mutual Information Analyzer. We pick out $100$ most informative features (excluding the redundant ones), and we apply Mutual Information Analyzer to obtain $6$ clusters. Figure \ref{fig:mi_by_category} shows how many features each category contributes, and which cluster the feature belongs to.

We find that redundant features are pervasive among the highly informative features. We look at $183$ most informative features, and $45.36\%$ of them are redundant. This phenomenon suggests future feature set engineering may be able to find many redundant features to prune without hurting its performance for website fingerprints.   

Figure \ref{fig:mi_by_category} shows a cluster may consist of features from different categories. For example, Cluster2 has features from category 1, 8, and 14, and Cluster3 has features from category 1, 3, and 14. This phenomenon shows features from different categories may share much information (that's why they are clustered together). Figure \ref{fig:mi_by_category} also shows features from same category are not necessarily in the same cluster. For instance, the category 4 features are clustered into three different clusters. 

Figure \ref{fig:mi_by_category} also shows that categories do not necessarily have features to be included in clusters. We find that some categories lack top informative features, ending up with absence of their features in clusters. Here, we clarify that we don't claim WeFDE to be free of information loss. In fact, just like other dimension reduction approaches such as PCA, there is information loss in WeFDE, but it is minimal~\cite{kfingerprint}. It's also worth noting that though some categories or features are not chosen by WeFDE, this doesn't necessarily mean all of their information is lost, as their information may be shared and represented by other included categories or features.


Looking at $83$ redundant features, we find $33$ of them are redundant with total packet count. These features include incoming packet count and 2-gram (-1,-1), but exclude outgoing packet count (NMI between total and outgoing packet count is $0.4414$). The reason is that the number of incoming packets are much more than the number of outgoing packets in website browsing, so that total packet count is highly dependent on incoming packet count.

Due to page limit, we release all our measurement results in our GitHub repository.

\begin{figure}[!th]
\centering
\includegraphics[width=0.35\textwidth]{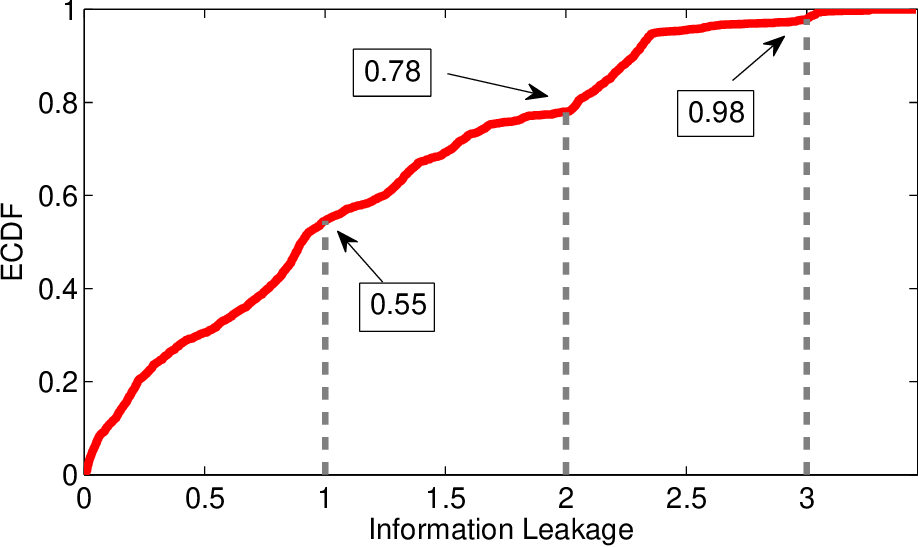}
\vspace{-5pt}
\caption{Information Leakage of Individual Features: {\normalfont Empirical Cumulative Distribution Function (ECDF) in the closed-world setting.}}
\label{fig:ecdf}
\vspace{-10pt}
\end{figure}

\begin{figure*}[!th]
\centering
\includegraphics[width=0.75\textwidth]{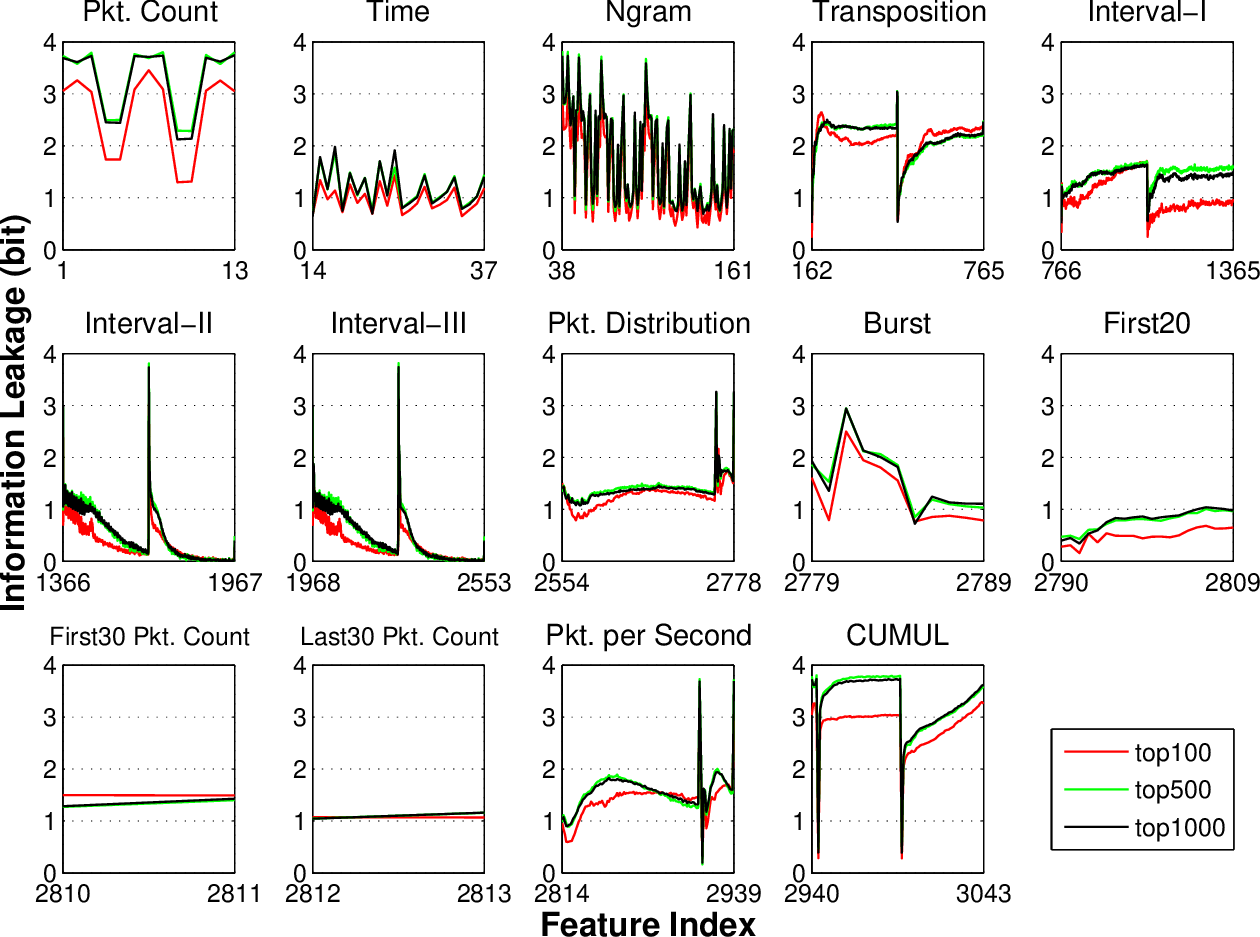}
\vspace{-5pt}
\caption{Closed-World Setting: Information Leakage for Individual Features (bit). {\normalfont Individual features have similar leakage with world size of 500 and 1000, indicating likely maximums}}
\label{fig:close_individual}
\vspace{-10pt}
\end{figure*}

\vspace{-6pt}
\section{Closed-World Information Leakage} \label{sec:closed}

In closed-world setting, an attacker is assumed to know the possible websites that a user may visit. The information leakage under this setting comes to \emph{which website is visited}. Appendix \ref{sec:entropy} gives more details about how to calculate this information. In the measurement, 
We adopt Alexa top 100 websites with $55779$ visits in our closed-world setting, as is shown in Table \ref{tab:dataset}. We assume equal prior probability for websites, and we set Monte Carlo sample number to $5000$. We measure $3043$ features' individual information leakage and their joint leakage by categories. We run this measurement and the following ones of this paper on a workstation with 32 cores (Intel Xeon CPU E5-2630 v3 @ 2.40GHz). The measurement time differs depending on the settings, but a typical measurement like the setting here can be finished within 10 hours. The following introduces part of our results. Full measurement results can be found at our anonymous GitHub repository.

{\bf Individual Information Leakage.}
Our measurement results upon individual features are shown in Figure~\ref{fig:ecdf}. Among these $3043$ features, we find: (a) $2.1\%$ features leak more than 3 bits information, meaning that an attacker is able to narrow down the possibilities to one eighth by any of these features; (b) $19.91\%$ features leak less than 3 bit but more than 2 bits information; (c) $23.43\%$ features leak 1 bit to 2 bits information; and (d) $54.55\%$ features leak less than 1 bit information. It is clear that nearly half of the features are able to help an attacker to reduce the size of the anonymity set by half. Yet our experiment shows that a single feature leaks no more than $3.45$ bits information, which is the maximum leakage we observe in our experiment from the feature of rounded outgoing packet count. We also observe that outgoing packet count without rounding leaks $3.26$ bits information, $0.19$ bit less than the rounded one. We observe similar information leakage increase by rounding for total packet count and incoming packet count. Our results confirm the observation in~\cite{wpes11-panchenko} that rounding packet count can help website fingerprinting attacks.

Web-browsing is characterized by asymmetric traffic. The incoming packets, which contain the requested contents, usually outnumber the outgoing packets carrying the request. A natural question is, does download stream having more packets leak more information than upload stream? The answer is no: download stream leaks $3.04$ bits information, $0.22$ bit less than the incoming stream. Our measurement suggests that defense design should give no less (if not more) attention to upload stream in hiding both streams' packet count.

The most informative timing feature is the average of inter-packet timing for download stream with $1.43$ bit leakage. Among inter-packet timing features, the maximum leaks the least information with around $0.7$ bit leakage. Another category of timing features is transmission time. We observe that the information leakage increases from $25\%$ percentile to $100\%$ percentile transmission time. The most information leakage for transmission time comes to the total transmission time, which leaks $1.21$ bit information. Furthermore, our measurement shows that information leakage of timing features has little difference for upload and download stream.

We also experiment the impact of world size on individual feature's information leakage. We try to answer: with a larger world size, whether the information leakage of indiviual features increases or decreases. We further adopt Alexa top 500 and top 1000 separately for closed-world setting, and we conduct the same information leakage measurement as above. Note that the information leakage upper bound under the world size 100, 500, and 1000 is $6.64$, $8.97$, and $9.97$ bits, respectively. 
Our finding is that the impact of world size on information leakage is minimal, as is shown in Figure~\ref{fig:close_individual}. Particularly, when the world size increases from 500 to 1000, the features' individual information leakage is almost the same. Further analysis will be given in Appendix \ref{sec:size}.

{\bf Joint Information Measurement.}  
Among the 100 most informative features, many of the features share redudant information with other features. We set a threshold to $0.9$, and if two features have mutual information larger than $0.9$, we would consider a feature sharing most of its information with another one. Our results show that 62 of the 100 most informative features can be represented by the other 38 features, demonstrating the prevalence of redundant features in website fingerprint. This finding shows the necessity and effectiveness of our Mutual Information Analyzer in recognizing features sharing redundant information. Figure~\ref{fig:closed_category} also shows that after including sufficient non-redundant features, the category information leakage tends to reach plateau. This phenomenon shows that we can approximate the information of a category by including sufficient non-redundant most informative features in this category.  



Categories such as Time, Ngram, Transposition, Interval-II, Interval-III, Packet Distribution, Packet per Second, and CUMUL leak most of the information about the visited websites; other categories such as Packet Count, Interval-I, Burst, First20, First30 Packet Count, Last30 Packet Count leak 5.75, 5.86, 6.2, 4.20, 1.29, and 1.03 bits information, respectively. Our measurement shows that Interval-II and Interval-III leak more information than Inerval-I, with 6.63 bits for both Interval-II and Interval-III. In addition, we find that Interval-II and Interval-III are faster than Interval-I in reaching the plateau, indicating the former twos not only leak more information but also with less features. It is clear that recording intervals by their frequency of packet count (adopted in Interval-II and Interval-III) is more preferable than recording them in sequence (Interval-I).

We also experiment the impact of world-size on information leakage upon categories in closed-world setting. We find that with the increase of world size, most categories exhibit more information leakage, except First30 and Last30 Packet Count. Note that categories such as First20, Burst, Packet Count show little increase when the world size increases from 1000 to 2000. We leave the discussion to Appendix \ref{sec:size}.

\begin{figure}[h]
\centering
\subfloat[]{\includegraphics[width=0.23\textwidth]{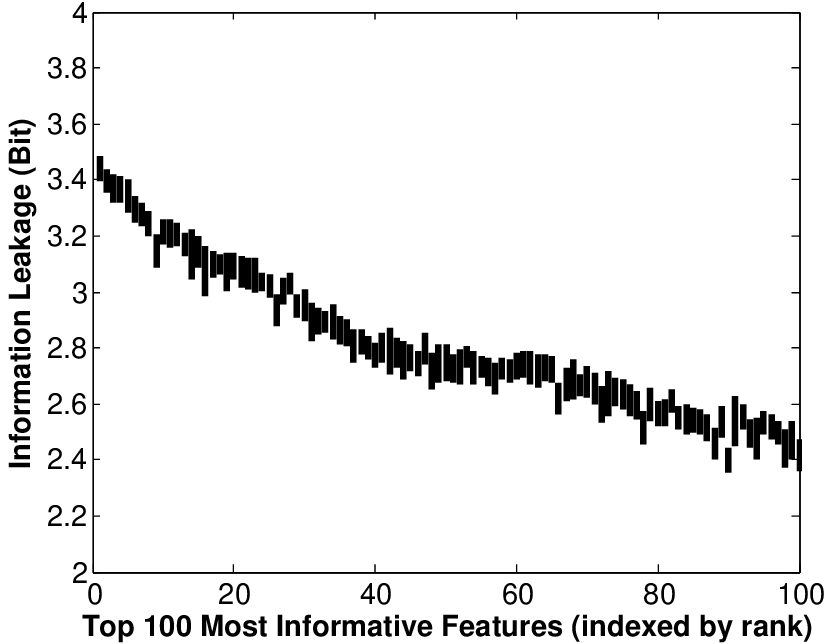}}
\subfloat[]{\includegraphics[width=0.23\textwidth]{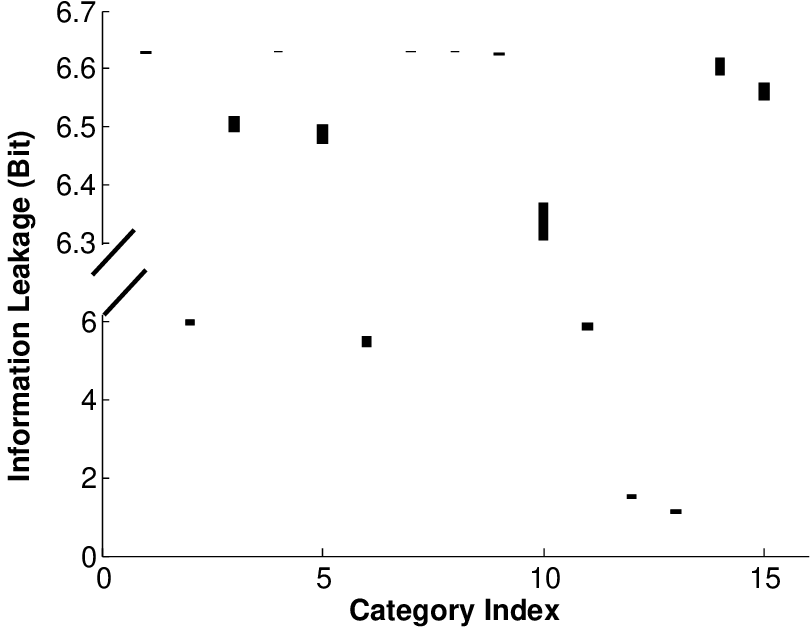}}
\vspace{-5pt}
\caption{Information Leakage Measurement Validation: {\normalfont $90\%$ Confidence Interval for the Measurement}}\label{fig:measure_conf}
\vspace{-10pt}
\end{figure}

\section{Validation} \label{sec:validation}
This section validates our measurement results by bootstrapping~\cite{efron1992bootstrap}. Bootstrapping is a statistical technique which uses random sampling with replacement to measure the properties of an estimator. More details about bootstrapping are given in Appendix \ref{app:bootstrap}.


{\bf Measurement Validation.}
This section shows how accurate our measurement is. We adopt bootstrapping with $20$ trials to give the $90\%$ confidence interval for the information leakage measurement. Figure \ref{fig:measure_conf} (a) shows the confidence intervals for top $100$ most informative features. We find that the width of the intervals is less than $0.178$ bit, and the median is around $0.1$ bit. Figure \ref{fig:measure_conf} (b) gives the $90\%$ confidence interval for $15$ categories. The width of these intervals is less than $0.245$ bit, with the median $0.03$ bit. We find that the interval of interval-I has the largest width. The bootstrapping results validate our information leakage measurement.

\begin{figure}[t]
  \centering
  \includegraphics[width=0.23\textwidth]{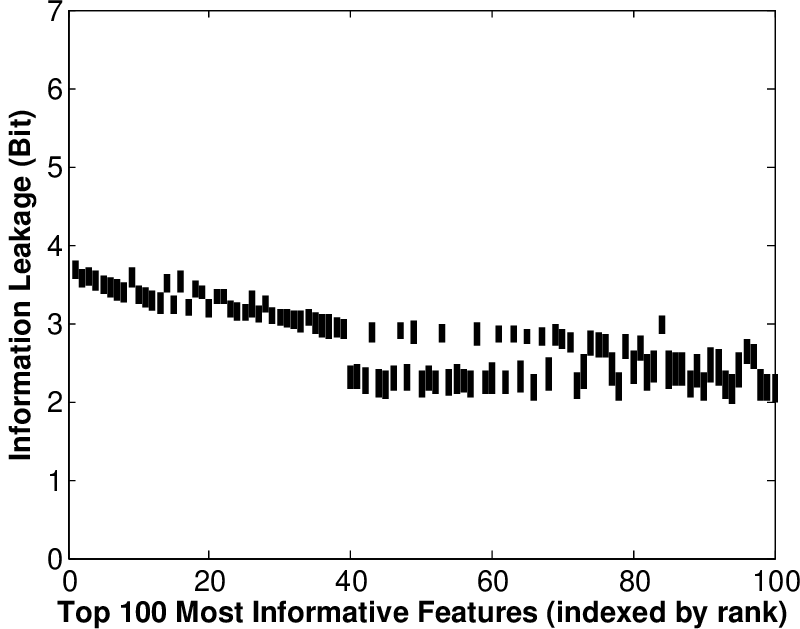}
  \includegraphics[width=0.23\textwidth]{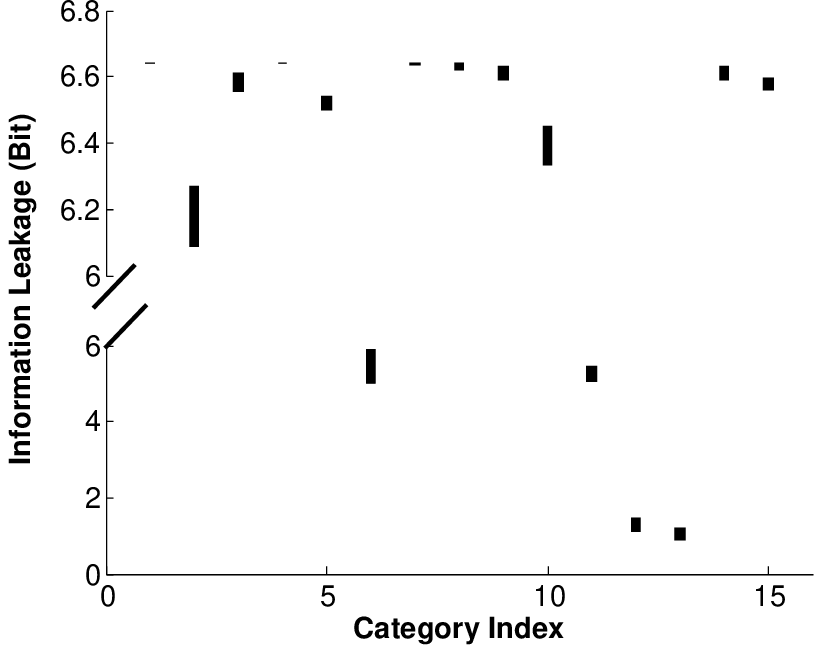}
\vspace{-5pt}
  \caption{Dataset and Generalization: {\normalfont the $90\%$ confidence interval by bootstrapping}}
  \label{fig:website_subsampling}
\vspace{-10pt}
\end{figure}

{\bf Dataset Validation.}
We use the top $100$ Alexa websites in the closed-world setting, as do previous works. But what if the top $100$ Alexa websites are not representative for Tor networks? Do our information leakage results still hold? While the representative websites are still unknown, we are able to validate our results by bootstrapping.

In the experiment, we have $2200$ websites for bootstrapping. In each round, we randomly sample $100$ websites {\em without} replacement to construct the bootstrapped dataset. Here we didn't use sampling with replacement because it makes less sense that the same website is included twice in a bootstrapped dataset. This special bootstrapping technique is also called subsampling~\cite{subsampling}. Repeating the same procedure $n$ times ($n=20$ in our experiment), we have $n$ such datasets to obtain $n$ bootstrapped measurements. Finally, we get the bootstrapped confidence interval for validation.

Figure \ref{fig:website_subsampling} displays the $90\%$ confidence interval for the top $100$ most informative features and $15$ categories of features. Not surprisingly, including different websites in the closed-world setting does make a difference in the measurement, but Figure \ref{fig:website_subsampling} shows such impact is very limited. Among top $100$ informative features, most of them have confidence interval with less than $0.5$ bit width, so do most of categories (even less for some categories). The exception only comes to category Interval-I. By bootstrapping, we validate our information leakage results even when the true representative websites are still unknown.

\vspace{-6pt}
\section{Information Leakage in WF Defenses}\label{sec:info_acc}

This section firstly gives the theoretical analysis on why accuracy is not a reliable metric to validate a WF defense. Then we measure the WF defenses' information leakage to confirm the analysis. Note that we choose the closed-world setting in the evaluation, as the setting is most advantageous for attackers, and we can get an upper bound for the defense's security.

\vspace{-6pt}
\subsection{Accuracy and Information Leakage}

The popular method to tell whether a WF defense is secure or not is to look at the classification accuracy under different WF attacks. If the defense is able to achieve low classification accuracy, it is believed to be secure. Here, we raise the question: does low classification accuracy always mean low information leakage? This question matters because if not, low classification accuracy would not be sufficient to validate a WF defense. To answer this question, we analyze the
relation between
information leakage and accuracy. We find that given a specific accuracy, the actual information leakage is far from certain.


\textbf{Theorem 1.} Let $\{c_1, c_2, \cdots, c_n\}$ denote a set of websites with prior probabilities $p_1, p_2, \cdots, p_n$, and $v_i$ denote a visit to website $c_i$. Suppose a website fingerprinting classifier $D$ which recognizes a visit $v_i$ to be $D(v_i)$. The classifier would succeed if $D(v_i) = c_i$, otherwise it fails. Assume a defense has been applied, and this classifier has $\alpha$ accuracy in classifying each website's visits. Then the information leakage obtained by the classifier is uncertain: the range of the possible information leakage is 
\begin{equation}
(1-\alpha)log_2(n-1)
\end{equation}

\emph{Proof:} see Appendix \ref{app:theorem1}
%
%
%
%

The reason for such uncertainty is that classification accuracy is "all-or-nothing". The classifier just makes one trial, and accuracy counts a miss as failure. But even a miss doesn't necessarily mean no information leakage. It is possible that the attacker could make a hit with the second or third trials, which indicates high information leakage; it also possible that the attacker could not do so without many trials, which equals to low information leakage. Accuracy alone cannot tell which case is true, as a result, the information leakage for a given accuracy is uncertain.

\medskip

\noindent \textbf{An Example.} Figure~\ref{fig:acc_vs_inforange} shows an example for the theorem. Note that the range is invariable no matter what we assume for websites' prior probability. We can see a wide range of possible information leakage when a low accuracy is given, showing that low accuracy doesn't necessarily guarantee low information leakage.


\begin{figure}[t]
  \centering
  \includegraphics[width=0.38\textwidth]{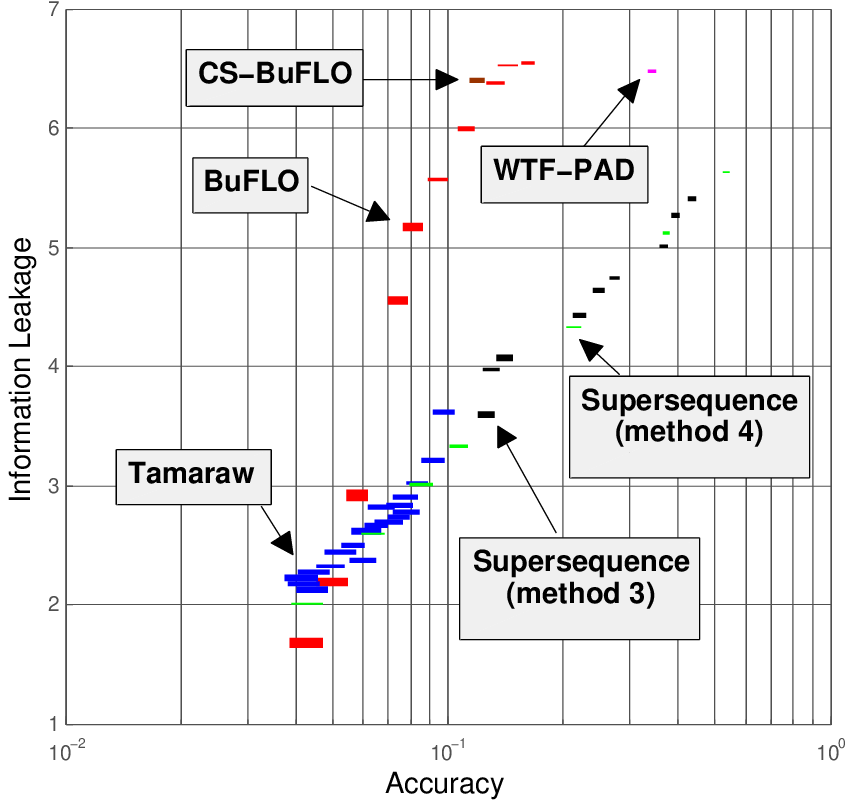}
  \vspace{-5pt}
  \caption{Website Fingerprinting Defenses: Accuracy vs. Information Leakage. {\normalfont Upon each type of defensed traces, we evaluate the overall information leakage and the classification accuracy at the same time. The results demonstrate the discrepancy between accuracy and information leakage}}
  \label{fig:acc_vs_info}
  \vspace{-10pt}
\end{figure}

\subsection{Measurement Results for WF defenses}

We include Tamaraw~\cite{cai2014}, BuFLO~\cite{buflo}, Supersequence~\cite{wang2014effective}, WTF-PAD~\cite{juarez2016toward}, and CS-BuFLO~\cite{wpes14-csbuflo} to quantify the information leakage upon defensed traffic. 

We adopt the implementation of BuFLO, Tamaraw, and Supersequence~\cite{wang_homepage} to generate the defensed traffic, with $\tau= $5, 10, 20, 30, 40, 50, 60, 80, 100, or 120 for BuFLO, $L$ ranging from $10$ to $100$ with step $10$ and from $200$ to $1000$ with step $100$ for Tamaraw. 
We include the method 3 of Supersequence, with 2, 5, or 10 super clusters, and 4, 8, or 12 stopping points. 
We also include the method 4 of Supersequence, with 2 super clusters, 4 stopping points, and 2, 4, 6, 8, 10, 20, 35, or 50 clusters. 
We use the implementation~\cite{marc_github} to create the WTF-PAD traffic. We were recommended to use the default \texttt{normal\_rcv} distributions on our dataset, as finding an optimal set of distributions for a new dataset is currently a work in progress~\cite{marc_github}. We apply the KNN classifier~\cite{wang2014effective} on our WTF-PAD traces, and we can get similar accuracy ($18.03\%$ in our case). This classification result validates our WTF-PAD traces. We use the implementation~\cite{naive_not_bayes} to generate simulated CS-BuFLO traces.  

Upon each type of defensed traces, we evaluate the overall information leakage and the classification accuracy at the same time. 
The measurement is conducted in closed-world setting with $94$ websites. To evaluate the total information leakage, we assume equal prior probability for websites and adopt $k=5000$ for Monte Carlo Evaluation. We use bootstrapping with $50$ trials to estimate the $96\%$ confidence interval for the information leakage and accuracy. For details about bootstrapping, please see Appendix \ref{app:bootstrap}. Note that we redo the dimension reductions for each defense, as a WF defense changes the information leakage of a feature and the mutual information between any two features.
The classifier we adopt is a variant of the KNN classifier~\cite{wang2014effective}. 
The only change we make is the feature set: we use our own feature set instead of its original one. The purpose is to have equivalent feature sets for classifications and information leakage measurements.
The reason for choosing this KNN classifier is that it is one of the most popular website fingerprinting classifiers to launch attacks and evaluate defense mechanisms.
It's also worth noting that the original feature set of the KNN classifier is a subset of our feature set. The experimental results are shown in Figure~\ref{fig:acc_vs_info}.


\vspace{-6pt}
\subsection{Accuracy is inaccurate}

Accuracy is widely used to compare the security of different defenses. A defense mechanism is designed and tuned to satisfy a lower accuracy as an evidence of superiority over existing defenses~\cite{buflo}. With defense overhead being considered, new defense mechanisms~\cite{cai2014,juarez2016toward} are sought and configured to lower the overhead without sacrificing accuracy too much. But if accuracy fails to be a reliable metric for security, it would become a pitfall and mislead the design and deployment of defense mechanisms. This section describes the flaws of accuracy and proves such a possibility.

{\bf Accuracy may fail because of its dependence on specific classifiers.} If a defense achieves low classification accuracy, it's not safe to conclude that this defense is secure, since the used classifiers may not be optimal. More powerful classifiers may exist and output higher classification accuracy. We prove this possibility in our experiment. To validate WTF-PAD, four classifiers were used including the original KNN classifier, and the reported highest accuracy was $26\%$. But using the KNN classifier with our feature set, we observe $33.99\%$ accuracy. Recent work~\cite{2018arXiv180102265S} even achieves $90\%$ accuracy against WTF-PAD. This work also confirms our measurement that the information leakage of WTF-PAD is high ($6.4$ bits), indicating that WTF-PAD is not secure.
Thus, accuracy is not reliable to validate a defense because of its dependence on specific classifiers.


{\bf Defenses having equivalent accuracy may leak varying amount of information.} Figure~\ref{fig:acc_vs_info} demonstrates such a phenomenon when taking BuFLO ($\tau=40$) and Tamaraw ($L=10$) into consideration. Accuracy of both defenses is nearly equivalent, with $9.39\%$ for BuFLO and $9.68\%$ for Tamaraw. In sense of accuracy, BuFLO ($\tau=40$) was considered to be as secure as Tamaraw($L=10$). However, our experimental results disapprove such a conclusion, showing BuFLO ($\tau=40$) leaks $2.31$ bits more information than Tamaraw (which leaks $3.26$ bits information). We observe the similar phenomenon between WTF-PAD and Supersequence. 

{\bf More importantly, a defense believed to be more secure by accuracy may leak more information.} Take BuFLO ($\tau=60$) as an example. Its accuracy is $7.39\%$, while accuracy of Tamaraw with $L=10,20,30$ is $9.68\%$, $9.15\%$, and $8.35\%$ respectively. Accuracy supports BuFLO ($\tau=60$) is more secure than Tamaraw with $L=10,20,30$. However, our measurement shows that  BuFLO ($\tau=60$) leaks $4.56$ bit information, $1.3$ bit, $1.61$ bit, and $1.75$ bit more than Tamaraw with $L=10, 20, 30$! Take WTF-PAD as another example. The accuracy for WTF-PAD is $33.99\%$, much lower than the $53.19\%$ accuracy of Supersequence method 4 with 2 super clusters, 50 clusters, and 4 stopping points. But the information leakage of WTF-PAD is around $6.4$ bits, much higher than the leakage of the latter which is about $5.6$ bits. Our experimental results prove the unreliability of accuracy in comparing defenses by security.



\begin{figure*}[t]
\centering
\includegraphics[width=0.82\textwidth]{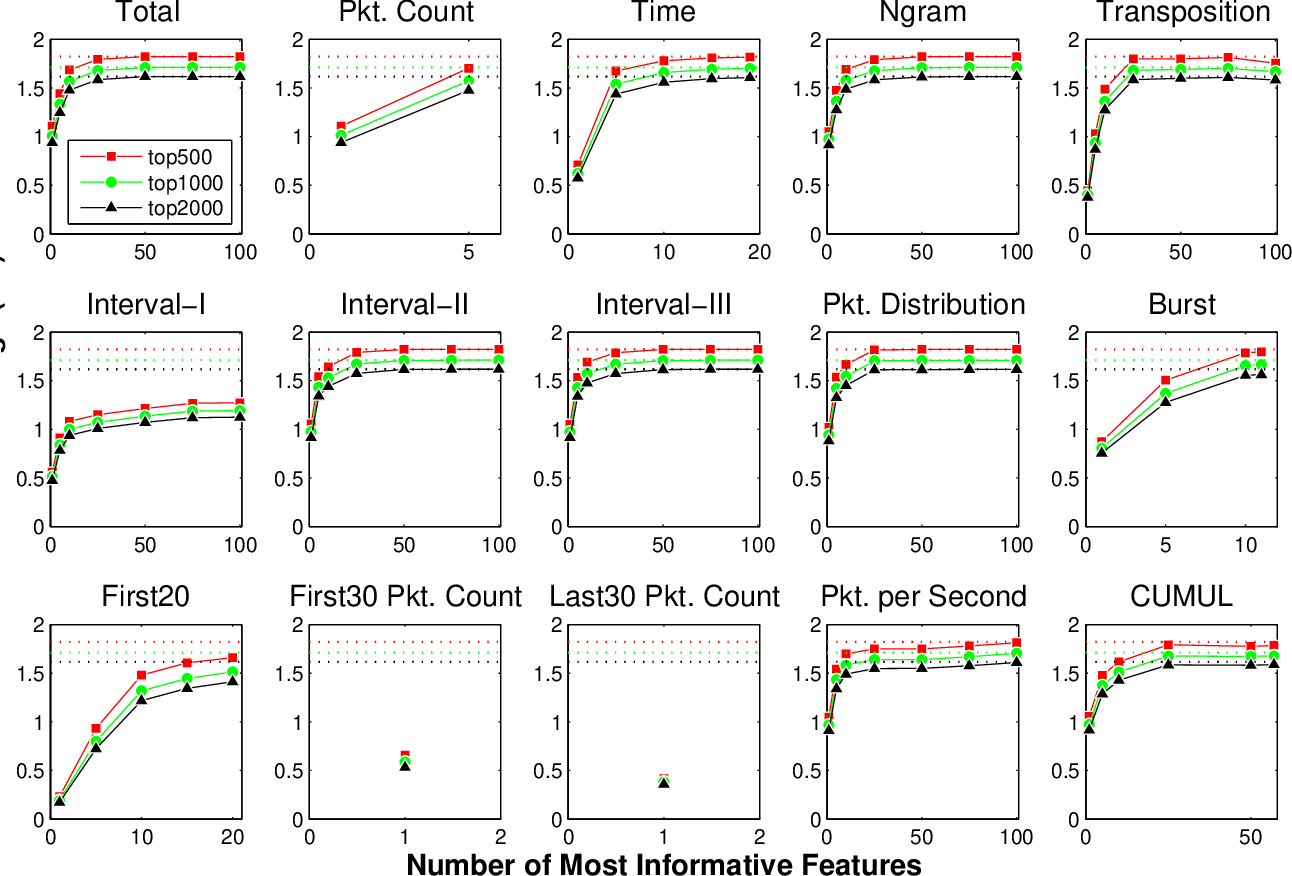}
\vspace{-5pt}
\caption{Open-World Setting: Information Leakage by Categories. {\normalfont This figure shows that the difference in world setting has little impact on categories' capability in leaking information}}
\label{fig:openw_category}
\vspace{-10pt}
\end{figure*}

\vspace{-6pt}
\section{Open-world Information Leakage} \label{sec:open}

In the closed-world scenario, the attacker knows all possible websites
that a user may visit, and the goal is to decide \emph{which} website
is visited; In the open-world setting, the attacker has a set of
monitored websites and tries to decide whether the monitored websites are visited and which one. The difference in information leakage is that the
open-world has $n + 1$ possible outcomes, whereas the closed-world
has $n$ outcomes where $n$ is the number of (monitored) websites. We include the details about how to quantify this information in Appendix \ref{sec:entropy}. The following describes part of our results for the open-world information leakage. For more information, please visit our GitHub repository. 

\textbf{Experiment Setup.} We adopt the list of monitored websites from \cite{wang2014effective} and collected $17984$ traffic instances in total. Our non-monitored websites come from Alexa's top 2000 with $137455$ instances in total. We approximate the websites' prior probability by Zipf law~\cite{adamic2002zipf, dns_tor}, which enables us to estimate a website's prior probability by its rank. We conduct experiments with top $500, 1000, 2000$ non-monitored websites separately, and we show the experimental results in Figure~\ref{fig:openw_category}.

Figure~\ref{fig:openw_category} shows that the open-world information leakage is decreased when including more non-monitored websites,  with $1.82, 1.71, 1.62$ bit for top500, top1000, top2000, respectively. Including more non-monitored websites decreases the entropy of the open-world setting rather than increasing it. The reduced information is in part because of the prior on monitored websites.
Compared with closed-world setting with similar world size, open-world scenario carries much less information.

Similar with the closed-world setting, Figure~\ref{fig:openw_category} shows that most categories except First20, First30 and Last30 Packet count, and Interval-I leak most of the information. This shows that the difference in world setting has little impact on categories' capability in leaking information.

\begin{figure}[t]
\centering
\subfloat[]{\includegraphics[width=0.23\textwidth]{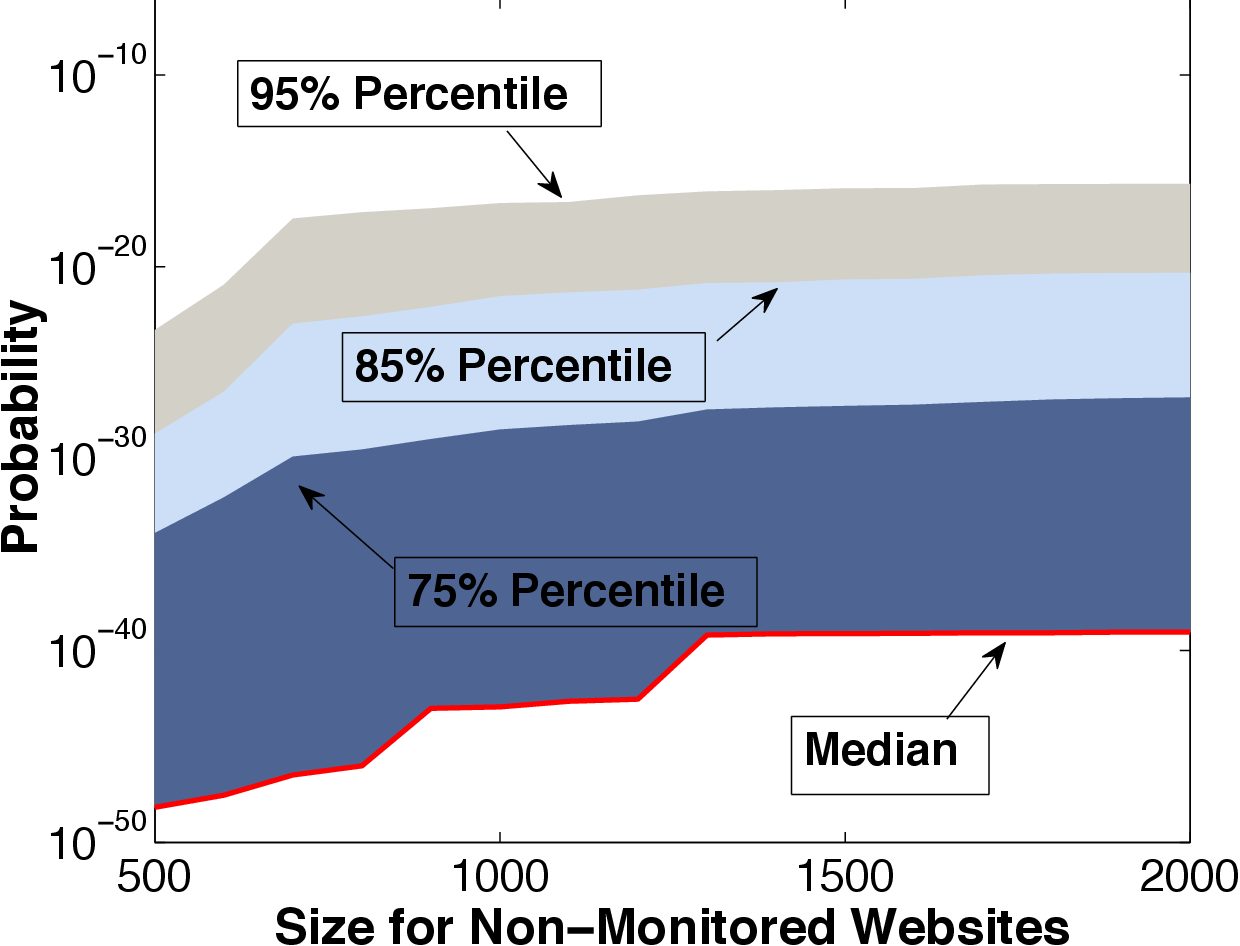}}
\subfloat[]{\includegraphics[width=0.23\textwidth]{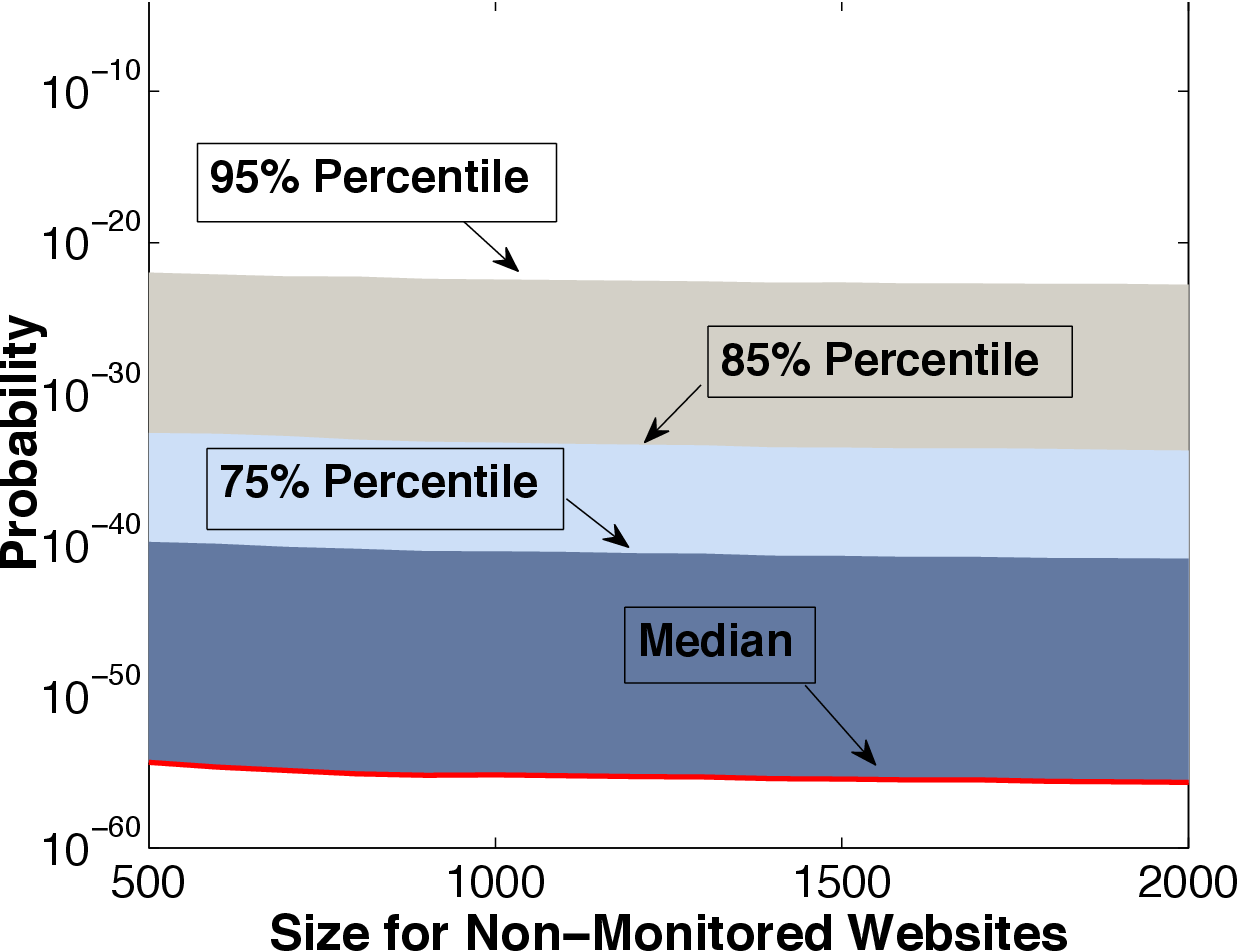}}
\caption{Size of the Non-Monitored Websites and Open-World Information Leakage Measurement: \normalfont (a) Monitored Samples at Non-Monitored AKDE, and (b) Non-Monitored Samples at Monitored AKDE}
\label{fig:openw_size}
\end{figure}

We also investigate how the size of the non-monitored websites influences our measurement. We focus on the total leakage and build the AKDE models for the non-monitored websites with the varying size of the non-monitored, respectively. We evaluate how the difference of these AKDE models influences measurement. Specifically, we evaluate (a) how monitored samples are evaluated at these AKDE models, and (b) how samples generated by these AKDE models are evaluated at the monitored AKDE. Figure~\ref{fig:openw_size} shows the results. Figure~\ref{fig:openw_size} (a) shows that these AKDE models of the non-monitored, though differing in size, assign low probability (below $10^{-10}$ with $95\%$ percentile) to monitored samples. Figure~\ref{fig:openw_size} (b) shows that though these AKDE models for the non-monitored generate different samples, the difference on how these samples are evaluated by the AKDE model of the monitored is little: they are all assigned low probability below $10^{-20}$ with $95\%$ percentile. The results lead to the estimation that introducing extra lower rank websites into the non-monitored set would not significantly change the low probability that the non-monitored AKDE assigns to monitored samples, and the low probability that the monitored AKDE assigns to samples generated by the non-monitored AKDE, thanks to the low prior probability of these websites. The information leakage is therefore little impacted.


\vspace{-6pt}
\section{Discussion} \label{sec:discuss}

{\bf WF Defense Evaluation.} We have discussed why using accuracy alone to validate a WF defense is flawed. Note that we don't mean that WF defense designers should keep away from accuracy. In fact, accuracy is straightforward and easy to use, and it is suitable for the initial check on WF defense design: if the classification accuracy is high, then the defense design is not secure. But if the classification accuracy is low, it doesn't necessarily mean the defense
is secure. We recommend defense designers to include other methods to further validate the defense. A potential approach to use is top-k accuracy, which allows WF attackers to make $k$ guesses instead of one, and if the $k$ guesses contain the right one, then the attackers succeed, otherwise, they lose. Another approach is information leakage measurement tools such as WeFDE. WeFDE gives information-theoretic perspective in evaluating WF defenses. When evaluating a defense by a classifier, a test case unseen by the training dataset is likely to be misclassified. But we can imagine that enlarging the dataset would effectively avoid such misclassification. This issue favors the defense design, and it is more likely to happen in probabilistic defenses such as WTF-PAD. Using WeFDE to evaluate a defense doesn't have this problem, as all data points are used to build the model for information leakage measurement. In addition, WeFDE is independent from any classifier, and it avoids the flaw of accuracy. The comparison of these approaches in validating WF defenses is out of the scope of this
paper, and we leave it in our future work.   

{\bf WeFDE's Other Applications.}
WeFDE can be used to launch website fingerprinting attacks. WeFDE models the likelyhood function of website fingerprints, so that given a test case, WeFDE is able to decide the probability of the test case being a visit to each website. It could be further combined with prior information about likely destinations to draw Bayesian inference~\cite{Greschbach2016a}.

WeFDE can be used to bootstrap a defense design. WeFDE can tell defense designers the information leakage from features and categories, so that designers could be guided to work on specific highly informative features and categories for hiding. In addition, when defenses are designed for a individual server or client to adopt~\cite{cherubin2017website}, WeFDE could suggest popular fingerprints to emulate. 
In our future work, we will explore more about using WeFDE to bootstrap a defense design.

{\bf Limitations.}
One limitation of WeFDE is its dependence on the feature set. 
Though we try to include all known features to generalize WeFDE's results, unknown informative features may exist and not be included. Fortunately, as long as new features are discovered and reported by future studies, we can always update our feature set and re-evaluate the leakage.


\vspace{-6pt}
\section{Conclusion}\label{sec:conclusion}

We develop a methodology and tools that allow measurement of
the information leaked by a website fingerprint. This gives us a more
fine-grained analysis of WF defense mechanisms than the
``all-or-nothing'' approach based on evaluating specific classifiers. By measuring defenses' information leakage and their accuracy, we find that using classification accuracy to validate a defense is flawed.

\begin{acks}
We would like to thank Tao Wang, Marc Juarez, Michael Carl Tschantz, Vern Paxson, George Karypis, Sheng Chen for the helpful discussions which improved this paper. We thank Marc Juarez {\em et al.} for helping us on Tor Browser Crawler. Shuai specially thanks his wife Wen Xing for her support and valued encouragement in this work. This paper is supported by NSF 1314637 and NSF 1815757.
\end{acks}

\bibliographystyle{ACM-Reference-Format}
\bibliography{mybib}

\appendix

\section{bootstrapping: accuracy estimation for information leakage quantification} \label{app:bootstrap}

We use bootstrapping~\cite{efron1992bootstrap} to estimate the accuracy of our information-theoretic measurement. bootstrapping is a statistical technique which uses random sampling with replacement to measure the properties of an estimator.



We implement bootstrapping to estimate the confidence interval of the information leakage. We describe our bootstrapping in the following:
\begin{itemize}
\item Step 1: for the observations of the each website, we apply random sampling with replacement, in which every observation is equally likely to be drawed and is allowed to be drawed more than once (with replacement). We let the sampling size be equal to observation size. 
\item Step 2: we apply our measurement on the newly constructed dataset of resamples and obtain the information leakage. 
\item Step 3: Step 1 and Step 2 are repeated $K$ times to obtain $K$ values for the information leakage; We therefore find the $\texttt{CI}$ confidence interval based on these $K$ values.   
\end{itemize}

Subsampling~\cite{subsampling} is a special bootstrapping technique. It uses sampling without replacement, and its sampling size is usually much smaller than the observation size.


\section{Proof of Theorem 1}\label{app:theorem1}
Let $I(D;V)$ denote the information leakage that the classifier attains. we have 

\begin{equation}
\begin{split}
 I(D;V) =& H(D) - H(D|V) \\
    =& H(D) - \sum_{v_i\in V}{p(v_i)H(D|v_i)}
\end{split}
\end{equation} 

We then evaluate $H(D|v_i), v_i\in V$. With the accuracy $\alpha$, we have $Pr(D = v_i|v_i) = \alpha$. However, it is uncertain about the probability $Pr(D=v_j|v_i)$ where $j\neq i$, from the knowledge of the accuracy. We put the possibility in two extremes to obtain the range of possible evaluation. In one case, suppose that the classifier determines $v_i$ to be from the website $C_{\hat{j}}$ with probability $1-\alpha$, so that the maximum of $I(D;V)$ is obtained as 
\begin{equation}
\max\{ I(D;V) \} = H(D) + \alpha\log_2{\alpha} + (1-\alpha)\log_2{(1-\alpha)} 
\end{equation}

In the other case, suppose that the probabilities $Pr(D = v_j|v_i) = (1-\alpha)/(n-1)$ where $j\neq i$, which means that except the correct decision, the classifier determines the visit $v_i$ to belong to any website other than $C_i$ with equal probability. Such a case yields the minimum possible information leakage, which is:
 \begin{equation}
\min\{ I(D;V) \} = H(D) + \alpha\log_2{\alpha} +  (1-\alpha)log_2{\frac{1-\alpha}{n-1}} 
\end{equation}

As a result, the range of potential information leakage $I(D;V)$ conveyed by the accuracy $\alpha$ is 
\begin{equation}
(1-\alpha)log_2(n-1)
\end{equation}

\section{Adaptive Kernel Density Estimate in WeFDE} \label{sec:kde}
This section gives details about Adaptive Kernel Density Estimate (AKDE) and the bandwidth selection approaches in WeFDE. 

We start by how WeFDE applies AKDE to estimate a single feature's probability distribution:
\begin{equation}
\hat{p}_(\bar{f}|c_j) = \frac{1}{n} \sum_{c=1}^{m}{\frac{1}{h_c} K(\frac{\bar{f}-p_c}{h_c}) }\
\end{equation}
where

$h_c$ is the bandwidth in AKDE,

$K(\cdotp)$ is the kernel function of a Gaussian, and

$p_1,p_2, \cdots, p_m$ are the observations for $\bar{f}$ in visiting $c_j$

\medskip

Choosing proper bandwidths is important for AKDE to make an accurate estimate. If a feature is continuous, WeFDE adopts plug-in estimator\cite{bw_survey}. In case of failure, we use the rule-of-thumb approach~\cite{bw_survey} as the alternative. If the feature is discrete, we let the bandwidth be a very small constant (0.001 in this paper). The choice of the small constant has no impact on the measurement, as long as each website uses the same constant as the bandwidth in its AKDE.  

Our AKDE may model a discrete feature as if it's continuous. The reason is that the domain of the feature $\bar{f}$ could be very large and requires much more samples than we can collect in order to accurately estimate in discrete. Take total packet count of Facebook.com as an example. We observe $238$ different values from our $600$ samples in the range of $576$ and $1865$. If the feature is processed as discrete, the estimated probability for observed values would be inaccurate, and the missed values would be considered impossible in the inappropriate way. Our solution is to consider such a discrete feature to be continuous, so that the kernels would smooth the probability distribution estimation and assign an appropriate probability to the missed values, making our measurement more accurate. 

Our AKDE is able to distinguish a continuous-like discrete feature. Take the feature of transmission time as an example. This feature is used to be continuous, but when defenses such as Tamaraw~\cite{cai2014} are applied, the feature would become discrete. Our AKDE is able to recognize by two approaches. The first approach is about using a threshold $\beta$. If the same traffic instances are observed more than $\beta$ times in our dataset, these instances are distinguished as discrete cases, and our AKDE would consider their features to be discrete. The second approach is template matching used in BuFLO case. We precompute a pattern of traffic believed to be discrete, and we consider the instances matching the pattern as discrete as well. In case of BuFLO, the pattern is the resulted traffic instance with transmission time $\tau$. 

Moreover, our AKDE can handle a feature which is partly continuous and partly discrete (or in other words, a mixture of continuous and discrete random variables). Such features exist in a WF defense such as BuFLO~\cite{buflo} which always sends at least T seconds. These features would be discrete if the genuine traffic can be completed within time $T$, otherwise, the features would be continuous. Thanks to AKDE which allows different observations to have their separate bandwidths, we compute the bandwidths separately for discrete and continuous feature values. According to~\cite{fruhwirth2006finite,mixture}, AKDE is able to model a feature with mixed nature by selecting adaptive bandwidths for its observations.

\section{Information Leakage in Two Worlds}\label{sec:entropy}
This section describes how to apply mutual information to quantify the information leakage in the closed-world setting and the open-world setting.

\medskip

\textbf{Closed-world Setting.} Suppose $C$ is a random variable denoting possible websites that a user may visit. Then the information leakage $I(F;C)$ in the closed-world scenario is:

\begin{equation} \label{eq:single_feature}
  \begin{split}
    I(C;F) &= H(C) - H(C|F) \\
           H(C)  &= -\sum_{c_i \in C}{\Pr(c_i)\log_2\Pr(c_i)} \\
    H(C|F) & =  \int_{\Phi} p(x) H(C|x) dx
  \end{split}
\end{equation}
where

$Pr(c_i)$ is the probability that the visited website is $c_i$,

$\Phi$ is the domain for the feature $F$, and

$p(x)$ is the probability density function for variable $x$.

\medskip

\textbf{Open-world Setting.} The information leakage $I(F;O)$ in the open-world scenario is:
\begin{equation}
I(F;O) = H(O) - H(O|F)
\end{equation}

\begin{equation}
\begin{split}
  H(O) &= - \sum_{c_i\in M}{ \Pr(c_i)\log_2\Pr(c_i) } \\
        &-\left\{\sum_{c_j\in N}{\Pr(c_j)}\right\}\log_2{\{\sum_{c_j\in N}{\Pr(c_j)}\}} 
\end{split}
\end{equation}

\begin{equation}
H(O|F) = \int_{\mathbb{F}} p(f) H(O|f) df
\end{equation}

\begin{equation}
\begin{split}
  H(O|f) &= -\sum_{c_i\in M}{\Pr(c_i|f)\log_2(\Pr(c_i|f))}\\
 &-\left\{\sum_{c_j\in N}{\Pr(c_j|f)}\right\}\log_2{\{\sum_{c_j\in N}{\Pr(c_j|f)}\}} 
\end{split}
\end{equation}

where $O$ is a random variable denoting the visited website belongs to the monitored or the non-monitored, and if it is monitored, which one. $M$ denotes the monitored set of websites, and $N$ denotes the non-monitored set of websites. $\mathbb{F}$ denotes the domain for feature $F$.

\section{Feature Set} \label{sec:appfeature}
The following lists the 14 categories of features which are included in the state-of-art attacks.

\medskip
\noindent{\textbf{1. Packet count.}} Counting the number of packets is found helpful for an attacker. Specifically, we include the following features based on packet count: (a) the total packet count, (b) the count of outgoing packets, (c) the count of incoming packets, (d) the ratio between the incoming packet count and that of the total, and (e) the ratio between the outgoing packet count and that of the total.     

\medskip
\noindent{\textbf{2. Time Statistics.}} Firstly, we look at the packet inter-arrival time for the total, incoming, and outgoing streams, individually. We extract the following statistics and add them into our feature set: (a) maximum, (b) mean, (c) standard deviation, and (d) the third quartile. Secondly, we embrace the features based on transmission time. We add the first, second, third quartile and total transmission time into our feature set.

\medskip
\noindent{\textbf{3--4. Packet Ordering.}} 
we explore the n-gram features which are widely adopted features extracting packet ordering. A $n$-gram is a contiguous sequence of $n$ packet lengths from a traffic sequence. Let's take 2-gram as an example. Suppose the traffic sequence is $\langle (l_1,t_1)$, $(l_2,t_2)$, $(l_3,t_3)$, $(l_4,t_4)\rangle$, then the 2-grams are $(l_1,l_2)$, $(l_2,l_3)$ and $(l_3,l_4)$. We consider the frequencies of each grams as features and we measure bigram, trigram, 4-gram, 5-gram, and 6-gram for comparison.

In addition, the number of packets transmitted before each successive incoming or outgoing packets also captures the ordering of the packets. We record such features by scanning the first 300 packets of the incoming and those of the outgoing respectively.

\medskip
\noindent{\textbf{5--7 and 9. Intervals and Bursts.}} We firstly adopt interval-based features to capture the traffic bursts. An interval is defined as a traffic window between a packet and the previous packet with the same direction. 

We use two approaches for interval-based features: Interval-I~\cite{wang2014effective} records the first $300$ intervals of incoming packets and those of the outgoing, Interval-II~\cite{shi2009fingerprinting} uses a vector $\mathbf{V}$ in which $\mathbf{V}(i)$ records the number of intervals with the packet number $i$. We use two vectors to count the incoming and outgoing intervals separately, and we fix the vectors' dimension to be $300$ (An interval having more than $300$ packets is counted as a interval with $300$ packets). We also apply grouping~\cite{wpes11-panchenko} on $\mathbf{V}$ to obtain extra features: $\sum_{i=3}^{5}{\mathbf{V}(i)}$, $\sum_{i=6}^{8}{\mathbf{V}(i)}$, and $\sum_{i=9}^{13}{\mathbf{V}(i)}$. We name this approach to be Interval-III.  

We also adopt ~\cite{wang2014effective}'s approach of counting the bursts for outgoing packets. A burst of outgoing packets is defined as a sequence of outgoing packets, in which there are no two adjacent incoming packets. We extract the packet number in each burst and use the maximum and the average as features. We also add the total burst number, as well as the number of bursts with more than $5$ packets, $10$ packets, and $20$ packets, respectively.   
 
\medskip
\noindent{\textbf{8. Packet Distribution.}} We divide the packet sequence into non-overlapping chunks of $30$ packets and count the number of outgoing packets in first $200$ chunks as features. We ignore the chunks after the $200$ chunks if any, and pad $0$s to have $200$ features in case of having less than $200$ chunks\cite{wang2014effective}.

We also apply the approaches in \cite{kfingerprint} to have additional features: (a) calculate the standard deviation, mean, median, and maximum of the $200$ features, and (b) split them into $20$ evenly sized subsets and sum each subset to be new features.

\medskip
\noindent{\textbf{10--12. First 30 and Last 30 Packets.}} We explore the information leakage from the first and last 30 packets. Particularly, we include first 20 packets as features, and we extract the packet count features (incoming packet count and outgoing packet count) from the first and last 30 packets, respectively.

\medskip
\noindent{\textbf{13. Packet count Per Second.}} We count the packet number in every second. To make the feature number fixed, we count the first $100$ seconds and pad $0$s if the transmission time is less than $100$ seconds. The standard deviation, mean, median, minimum, and maximum of these features are also included. 

We also include the alternative count of packets per second features\cite{kfingerprint}. We split the packet count per second features into $20$ evenly sized subsets and sum each subset to obtain the alternative features.  

\medskip
\noindent{\textbf{14. CUMUL Features.}}
Panchenko {\em et al.}~\cite{ndss16} introduce the CUMUL features. A cumulative representation is extracted from the packet trace, and $n$ features are derived by sampling the piecewise linear interpolant of the representation at $n$ equidistant points. We adopt such features with $n = 100$.

It's worth noting that ``packet'' here refers to a Tor cell packet. We extract our features based on the cell packet traces. 
In addition, in 2011 ~\cite{wpes11-panchenko} includes a feature named HTML marker, which counts the total size of incoming packets from the first outgoing packet and the next outgoing packet. Such summation was considered to be the size of the HTML document and therefore is informative. We find such claim is not accurate anymore, and we find no updated details of how to reproduce such a feature. As a result, we do not include this feature in our measurement.

\section{World Size and Information Leakage} \label{sec:size}
In this section, we discuss the impact of the world size on our information leakage measurement. 

We start with the closed-world setting. We observe that with the increase of the world size, the information leakage for most categories and the total increases as well, while the individual information leakage of features is little impacted (particularly when the world size increases from 1000 to 2000). To explain the conflicting observations, we highlight the notion of maximum possible information leakage of a setting. A feature (or a set of features) leaks no more information than the information that the setting has. For example, in our closed-world setting with 100 websites, the total information leakage is $6.63$ bits. But if we let the world size be 2, the total leakage is no more than $1$ bit, no matter how distinguishable the fingerprint is. Therefore we argue that the increased information leakage with larger world size for most categories and the total is because the website fingerprint has the ability to leak more information than the information that our closed-world settings have. This phenomenon leads to an interesting question: what is the maximum information leakage the website fingerprint is able to leak in a sufficiently larger world size, which we include in our future work.

For the features' individual information leakage, we observe that the leakage in each setting is much less than the information that these setting have, and that the world size has little impact on the measurement. We explain the reason for the little impact of the world size by the following theorem:

\textbf{Theorem 2.} Let's consider $x$ closed-world settings with equal world size $n$. Suppose a feature $F = \bar{f}$ has \emph{valid} information leakage of $I_1, I_2, \cdots, I_x$ in each closed-world setting. In the combined closed-world setting with $nx$ world size, the information leakage of $F = \bar{f}$ would be $\frac{I_1 + I_2 + \cdots + I_x}{x}$.

\textbf{Proof:} let's denote the information leakage in each closed-world setting to be:
\begin{equation}
  I_l = log_2(n) + \sum_{l\in \{1,\cdots,n\}}{q_l(i)log_2(q_l(i))}
\end{equation}
, where $q_l(i)$ is the probability of visiting the $ith$ website in the $lth$ closed-world setting conditioned on $F = \bar{f}$.

In the combined closed-world setting, the information leakage of $F = \bar{f}$ is
\begin{equation}
  \begin{split}
    & log_2(nx) + \sum_{l\in \{1,\cdots,x\}}{\{\sum_{i\in \{1,\cdots,n\}}{\frac{q_l(i)}{x}log_2(\frac{q_l(i)}{x})}\}} \\
    &= log_2(n) + \frac{1}{x} \sum_{l\in \{1,\cdots,x\}} {\sum_{i\in \{1,\cdots,n\}}{q_l(i)log_2(q_l(i))}}  \\
    &= \frac{I_1 + I_2 + \cdots + I_x}{x}
\end{split}
\end{equation}

%

\medskip

This theorem reveals the relation between world size and information leakage. With each closed-world setting including sufficient websites, the combined larger world size would have little impact on the information leakage.

\begin{figure}[t]
  \centering
  \includegraphics[width=0.35\textwidth]{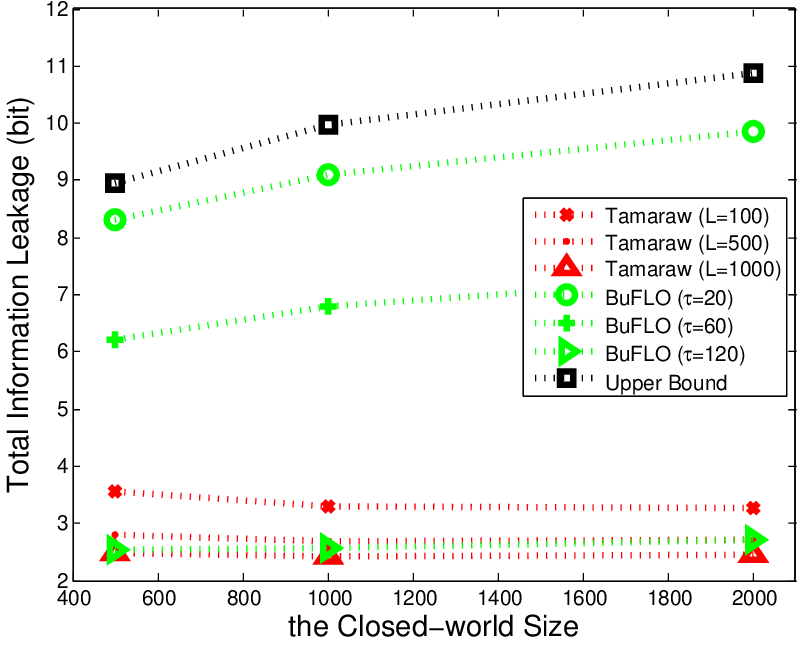}
  \vspace{-5pt}
  \caption{Defenses with Different World Size.}
  \label{fig:enlarge_defense}
  \vspace{-10pt}
\end{figure}


We also evaluate world size impact on defenses in closed-world setting. Figure~\ref{fig:enlarge_defense} shows that in Tamaraw, world size has little impact on information leakage. No matter how large the world size is, the information leakage for Tamaraw is around $3.3, 2.72, 2.45$ bits for $L=100, 500, 1000$. BuFLO with $\tau=120$ is not impacted by world size, but BuFLO with $\tau=20, 60$ see the increase of information leakage. The different impact from world size roots in BuFLO's mixed nature.

We discuss the world size impact on the open-world setting. Here the world size refers to the size of the non-monitored websites. We find that with a larger world size, the maximum information leakage decreases. In addition, as is shown in Section \ref{sec:open}, world size also has little impact on the measure.

\section{Monte Carlo Integral Evaluation}\label{sec:mc}
We use the Monte Carlo method~\cite{ghahramani2002bayesian} to evaluate the integral when measuring $H(C|\bar{f})$. Monte Carlo picks random points in the domain and uses these points to numerically approximate the definite integral:
\begin{equation}\label{eq:monte}
H(C|\bar{f}) \simeq \frac{1}{k} \sum_{i = 1}^{k} {H(C|\bar{f}^{(i)})}
\end{equation}

where

$\bar{f}^{(1)}, \bar{f}^{(2)}, \cdots, \bar{f}^{(k)}$ are the random samples, and 

$k$ is the size of the sample.

\medskip

Note that we apply importance sampling here, in which random samples are drawn from the distribution having probability density function $p(\bar{f})$. The sampling process is:
\begin{itemize}[noitemsep]
\item Step 1: decide sampling size. To accurately evaluate the integral by Monte Carlo method, sufficent samples are needed. In this paper, the total number of samples is set to be $k = 5000$. For each condition $c_j, j\in \{1, \cdots, n\}$, the number of samples is $k \cdot \Pr(c_j)$.
\item Step 2: sampling for different conditons. For each condition $c_j$, draw $k \cdot \Pr(c_j)$ samples from the distribution with conditional PDF $p(\bar{f}|c_j)$.
\end{itemize} 
We use this method to draw $k$ samples from the generic PDF $p(\bar{f})$.  We choose importance sampling over uniform  domain sampling (which would require a different estimation than Equation~\ref{eq:monte}) since it includes more samples for feature values that are more likely to happen. The benefit is that the ``important'' values in the integration are emphasized for higher precision.

\end{document}